\pdfoutput=1
\documentclass[12pt,a4paper]{article}
\usepackage{ifthen} % for conditional statements
\newboolean{pdflatex}
\setboolean{pdflatex}{true} % False for eps figures 

\newboolean{articletitles}
\setboolean{articletitles}{true} % False removes titles in references

\newboolean{uprightparticles}
\setboolean{uprightparticles}{false} %True for upright particle symbols

\def\paperauthors{LHCb collaboration} % Leave as is for PAPER, CONF and FIGURE
\def\paperasciititle{Measurement of CKM angle gamma in partially reconstructed B->D*h decays with D->Kshh} % Set ASCII title here !! MAKE sure it's only ASCII characters !! 
\def\papertitle{A model-independent measurement of the CKM angle \g in partially reconstructed $\Bpm \to \Dstar h^{\pm}$ decays with $\D \to \KS h^{+}h^{-}$ $(h=\pi, K)$} % Latex formatted title
\def\paperkeywords{{High Energy Physics}, {LHCb}} % Comma separated list
\def\papercopyright{\the\year\ CERN for the benefit of the LHCb collaboration} % new since 9/Apr/2018
\def\paperlicence{CC BY 4.0 licence}
\def\paperlicenceurl{https://creativecommons.org/licenses/by/4.0/}

\usepackage[top=1in, bottom=1.25in, left=1in, right=1in]{geometry}

\usepackage{microtype}
\usepackage{lineno}  % for line numbering during review
\usepackage{xspace} % To avoid problems with missing or double spaces after
\usepackage{caption} %these three command get the figure and table captions automatically small

\usepackage{graphicx}  % to include figures (can also use other packages)
\usepackage{color}
\usepackage{colortbl}
\graphicspath{{./figs/}} % Make Latex search fig subdir for figures
\usepackage{amsmath} % Adds a large collection of math symbols
\usepackage{amssymb}
\usepackage{amsfonts}
\usepackage{upgreek} % Adds in support for greek letters in roman typeset

\newcommand*\patchAmsMathEnvironmentForLineno[1]{%
\expandafter\let\csname old#1\expandafter\endcsname\csname #1\endcsname
\expandafter\let\csname oldend#1\expandafter\endcsname\csname
end#1\endcsname
 \renewenvironment{#1}%
   {\linenomath\csname old#1\endcsname}%
   {\csname oldend#1\endcsname\endlinenomath}%
}
\newcommand*\patchBothAmsMathEnvironmentsForLineno[1]{%
  \patchAmsMathEnvironmentForLineno{#1}%
  \patchAmsMathEnvironmentForLineno{#1*}%
}
\AtBeginDocument{%
\patchBothAmsMathEnvironmentsForLineno{equation}%
\patchBothAmsMathEnvironmentsForLineno{align}%
\patchBothAmsMathEnvironmentsForLineno{flalign}%
\patchBothAmsMathEnvironmentsForLineno{alignat}%
\patchBothAmsMathEnvironmentsForLineno{gather}%
\patchBothAmsMathEnvironmentsForLineno{multline}%
\patchBothAmsMathEnvironmentsForLineno{eqnarray}%
}

\usepackage{hyperxmp}

\usepackage[pdftex,
            pdfauthor={\paperauthors},
            pdftitle={\paperasciititle},
            pdfkeywords={\paperkeywords},
            pdfcopyright={Copyright (C) \papercopyright},
            pdflicenseurl={\paperlicenceurl}]{hyperref}
\usepackage[colorinlistoftodos,textsize=scriptsize]{todonotes}

\usepackage[bottom,flushmargin,hang,multiple]{footmisc}

\usepackage[all]{hypcap} % Internal hyperlinks to floats.

\usepackage{xspace} 
\usepackage{upgreek}

\def\lhcb   {\mbox{LHCb}\xspace}

\def\babar  {\mbox{BaBar}\xspace}
\def\belle  {\mbox{Belle}\xspace}

\def\besiii {\mbox{BESIII}\xspace}
\def\cleo   {\mbox{CLEO}\xspace}

\def\MagUp {\mbox{\em Mag\kern -0.05em Up}\xspace}

\ifthenelse{\boolean{uprightparticles}}%
{
 
 \def\Pgamma      {\ensuremath{\upgamma}\xspace}

 \def\Ppi         {\ensuremath{\uppi}\xspace}                 
                  
 \def\Prho        {\ensuremath{\uprho}\xspace}

 \def\Ppsi        {\ensuremath{\uppsi}\xspace}

 \def\PDelta      {\ensuremath{\Delta}\xspace}                 
 \def\PXi         {\ensuremath{\Xi}\xspace}                 
 \def\PLambda     {\ensuremath{\Lambda}\xspace}                 
 \def\PSigma      {\ensuremath{\Sigma}\xspace}                 
 \def\POmega      {\ensuremath{\Omega}\xspace}                 
 \def\PUpsilon    {\ensuremath{\Upsilon}\xspace}
 \let\oldPi\Pi
 \def\PPi         {\ensuremath{\oldPi}\xspace}

 \def\PB      {\ensuremath{\mathrm{B}}\xspace}                 
                  
 \def\PD      {\ensuremath{\mathrm{D}}\xspace}

 \def\PK      {\ensuremath{\mathrm{K}}\xspace}

 \def\Pb      {\ensuremath{\mathrm{b}}\xspace}                 
 \def\Pc      {\ensuremath{\mathrm{c}}\xspace}                 
 \def\Pd      {\ensuremath{\mathrm{d}}\xspace}

 \def\Ph      {\ensuremath{\mathrm{h}}\xspace}                 
 \def\Pi      {\ensuremath{\mathrm{i}}\xspace}

 \def\Ps      {\ensuremath{\mathrm{s}}\xspace}                 
                  
 \def\Pu      {\ensuremath{\mathrm{u}}\xspace}

 \def\thebaroffset{0.0em}
}
{
 
 \def\Pgamma      {\ensuremath{\gamma}\xspace}

 \def\Ppi         {\ensuremath{\pi}\xspace}                 
                  
 \def\Prho        {\ensuremath{\rho}\xspace}

 \def\Ppsi        {\ensuremath{\psi}\xspace}                 
                  
 \mathchardef\PDelta="7101
 \mathchardef\PXi="7104
 \mathchardef\PLambda="7103
 \mathchardef\PSigma="7106
 \mathchardef\POmega="710A
 \mathchardef\PUpsilon="7107
 \mathchardef\PPi="7105
                  
 \def\PB      {\ensuremath{B}\xspace}                 
                  
 \def\PD      {\ensuremath{D}\xspace}

 \def\PK      {\ensuremath{K}\xspace}

 \def\Pb      {\ensuremath{b}\xspace}                 
 \def\Pc      {\ensuremath{c}\xspace}                 
 \def\Pd      {\ensuremath{d}\xspace}

 \def\Ph      {\ensuremath{h}\xspace}                 
 \def\Pi      {\ensuremath{i}\xspace}

 \def\Ps      {\ensuremath{s}\xspace}                 
                  
 \def\Pu      {\ensuremath{u}\xspace}

 \def\thebaroffset{0.18em}
}
\newcommand{\offsetoverline}[2][\thebaroffset]{\kern #1\overline{\kern -#1 #2}}%

\makeatletter
\ifcase \@ptsize \relax% 10pt
  \newcommand{\miniscule}{\@setfontsize\miniscule{4}{5}}% \tiny: 5/6
\or% 11pt
  \newcommand{\miniscule}{\@setfontsize\miniscule{5}{6}}% \tiny: 6/7
\or% 12pt
  \newcommand{\miniscule}{\@setfontsize\miniscule{5}{6}}% \tiny: 6/7
\fi
\makeatother

\DeclareRobustCommand{\optbar}[1]{\shortstack{{\miniscule (\rule[.5ex]{1.25em}{.18mm})}
  \\ [-.7ex] $#1$}}

\def\g      {{\ensuremath{\Pgamma}}\xspace}

\def\uquark    {{\ensuremath{\Pu}}\xspace}

\def\dquark    {{\ensuremath{\Pd}}\xspace}

\def\squark    {{\ensuremath{\Ps}}\xspace}

\def\cquark    {{\ensuremath{\Pc}}\xspace}

\def\bquark    {{\ensuremath{\Pb}}\xspace}

\def\hadron {{\ensuremath{\Ph}}\xspace}
\def\pion   {{\ensuremath{\Ppi}}\xspace}
\def\piz    {{\ensuremath{\pion^0}}\xspace}
\def\pip    {{\ensuremath{\pion^+}}\xspace}
\def\pim    {{\ensuremath{\pion^-}}\xspace}
\def\pipm   {{\ensuremath{\pion^\pm}}\xspace}
\def\pimp   {{\ensuremath{\pion^\mp}}\xspace}

\def\rhomeson {{\ensuremath{\Prho}}\xspace}
\def\rhoz     {{\ensuremath{\rhomeson^0}}\xspace}

\def\rhopm    {{\ensuremath{\rhomeson^\pm}}\xspace}

\def\kaon    {{\ensuremath{\PK}}\xspace}
\def\KorKbar {\kern \thebaroffset\optbar{\kern -\thebaroffset \PK}{}\xspace}

\def\Kp      {{\ensuremath{\kaon^+}}\xspace}
\def\Km      {{\ensuremath{\kaon^-}}\xspace}
\def\Kpm     {{\ensuremath{\kaon^\pm}}\xspace}
\def\Kmp     {{\ensuremath{\kaon^\mp}}\xspace}
\def\KS      {{\ensuremath{\kaon^0_{\mathrm{S}}}}\xspace}

\def\Kstarz  {{\ensuremath{\kaon^{*0}}}\xspace}

\def\Kstarpm {{\ensuremath{\kaon^{*\pm}}}\xspace}

\def\Dbar    {{\ensuremath{\offsetoverline{\PD}}}\xspace}
\def\D       {{\ensuremath{\PD}}\xspace}
\def\Db      {{\ensuremath{\Dbar}}\xspace}
\def\DorDbar {\kern \thebaroffset\optbar{\kern -\thebaroffset \PD}\xspace}
\def\Dz      {{\ensuremath{\D^0}}\xspace}
\def\Dzb     {{\ensuremath{\Dbar{}^0}}\xspace}
\def\Dp      {{\ensuremath{\D^+}}\xspace}
\def\Dm      {{\ensuremath{\D^-}}\xspace}

\def\DpDm    {\ensuremath{\Dp {\kern -0.16em \Dm}}\xspace}
\def\Dstar   {{\ensuremath{\D^*}}\xspace}

\def\Dstarz  {{\ensuremath{\D^{*0}}}\xspace}
\def\Dstarzb {{\ensuremath{\Dbar{}^{*0}}}\xspace}

\def\Dstarm  {{\ensuremath{\D^{*-}}}\xspace}

\def\Dstarmp {{\ensuremath{\D^{*\mp}}}\xspace}

\def\B       {{\ensuremath{\PB}}\xspace}

\def\BorBbar {\kern \thebaroffset\optbar{\kern -\thebaroffset \PB}\xspace}
\def\Bz      {{\ensuremath{\B^0}}\xspace}

\def\Bd      {{\ensuremath{\B^0}}\xspace}

\def\BdorBdbar {\kern \thebaroffset\optbar{\kern -\thebaroffset \Bd}\xspace}
\def\Bu      {{\ensuremath{\B^+}}\xspace}
\def\Bub     {{\ensuremath{\B^-}}\xspace}
\def\Bp      {{\ensuremath{\Bu}}\xspace}
\def\Bm      {{\ensuremath{\Bub}}\xspace}
\def\Bpm     {{\ensuremath{\B^\pm}}\xspace}

\def\Bs      {{\ensuremath{\B^0_\squark}}\xspace}

\def\BsorBsbar {\kern \thebaroffset\optbar{\kern -\thebaroffset \Bs}\xspace}

\def\psiprpr  {{\ensuremath{\Ppsi(3770)}}\xspace}

\def\Y#1S{\ensuremath{\PUpsilon{(#1S)}}\xspace}

\def\Lz          {{\ensuremath{\PLambda}}\xspace}

\def\LorLbar     {\kern \thebaroffset\optbar{\kern -\thebaroffset \PLambda}\xspace}

\def\Lb           {{\ensuremath{\Lz^0_\bquark}}\xspace}

\def\to                 {\ensuremath{\rightarrow}\xspace}

\def\CP                {{\ensuremath{C\!P}}\xspace}

\def\Vud  {{\ensuremath{V_{\uquark\dquark}^{\phantom{\ast}}}}\xspace}
\def\Vcd  {{\ensuremath{V_{\cquark\dquark}^{\phantom{\ast}}}}\xspace}

\def\Vub  {{\ensuremath{V_{\uquark\bquark}^{\phantom{\ast}}}}\xspace}
\def\Vcb  {{\ensuremath{V_{\cquark\bquark}^{\phantom{\ast}}}}\xspace}

\def\AT#1     {\ensuremath{A_{\mathrm{T}}^{#1}}\xspace}           % 2

\def\C#1      {\ensuremath{\mathcal{C}_{#1}}\xspace}                       % 9
\def\Cp#1     {\ensuremath{\mathcal{C}_{#1}^{'}}\xspace}                    % 7
\def\Ceff#1   {\ensuremath{\mathcal{C}_{#1}^{\mathrm{(eff)}}}\xspace}        % 9  
\def\Cpeff#1  {\ensuremath{\mathcal{C}_{#1}^{'\mathrm{(eff)}}}\xspace}       % 7
\def\Ope#1    {\ensuremath{\mathcal{O}_{#1}}\xspace}                       % 2
\def\Opep#1   {\ensuremath{\mathcal{O}_{#1}^{'}}\xspace}                    % 7

\newcommand{\nospaceunit}[1]{\ensuremath{\text{#1}}}       
\newcommand{\aunit}[1]{\ensuremath{\text{\,#1}}}       
\newcommand{\tev}{\aunit{Te\kern -0.1em V}\xspace}
\newcommand{\gev}{\aunit{Ge\kern -0.1em V}\xspace}
\newcommand{\mev}{\aunit{Me\kern -0.1em V}\xspace}
\newcommand{\kev}{\aunit{ke\kern -0.1em V}\xspace}
\newcommand{\ev}{\aunit{e\kern -0.1em V}\xspace}
 
\newcommand{\mevc}{\ensuremath{\aunit{Me\kern -0.1em V\!/}c}\xspace}
\newcommand{\gevc}{\ensuremath{\aunit{Ge\kern -0.1em V\!/}c}\xspace}
\newcommand{\mevcc}{\ensuremath{\aunit{Me\kern -0.1em V\!/}c^2}\xspace}
\newcommand{\gevcc}{\ensuremath{\aunit{Ge\kern -0.1em V\!/}c^2}\xspace}
\def\mum  {\ensuremath{\,\upmu\nospaceunit{m}}\xspace}

\def\fb   {\ensuremath{\aunit{fb}}\xspace}
\def\invfb   {\ensuremath{\fb^{-1}}\xspace}

\newcommand{\chisq}{\ensuremath{\chi^2}\xspace}

\newcommand{\chisqip}{\ensuremath{\chi^2_{\text{IP}}}\xspace}

\def\gsim{{~\raise.15em\hbox{$>$}\kern-.85em
          \lower.35em\hbox{$\sim$}~}\xspace}
\def\lsim{{~\raise.15em\hbox{$<$}\kern-.85em
          \lower.35em\hbox{$\sim$}~}\xspace}

\def\pt         {\ensuremath{p_{\mathrm{T}}}\xspace}

\def\ptot       {\ensuremath{p}\xspace}

\def\degrees{\ensuremath{^{\circ}}\xspace}

\def\evtgen     {\mbox{\textsc{EvtGen}}\xspace}

\def\geant      {\mbox{\textsc{Geant4}}\xspace}

\def\photos     {\mbox{\textsc{Photos}}\xspace}

\def\pythia     {\mbox{\textsc{Pythia}}\xspace}

\def\tell1  {TELL1\xspace}
\def\ukl1   {UKL1\xspace}

\newcommand{\ie}{\mbox{\itshape i.e.}\xspace}

\newcommand{\lhcborcid}[1]{\href{https://orcid.org/#1}{\hspace*{0.1em}\raisebox{-0.45ex}{\includegraphics[width=1em]{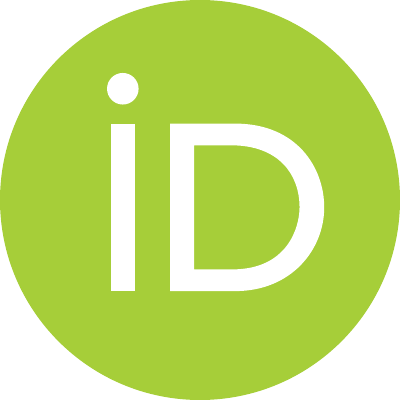}}}}

\usepackage{cite} % Allows for ranges in citations
\usepackage{mciteplus}
\usepackage{longtable} % only for template; not usually to be used in PAPERs
\usepackage{lscape}
\usepackage{adjustbox}

\usepackage[detect-all]{siunitx}
\DeclareSIUnit\eVperc{\eV\per\clight}
\DeclareSIUnit\clight{\text{\ensuremath{c}}}
\usepackage{subcaption}

\begin{document}

%%%%%%%%%%%%%%%%%%%%%%%%%
%%%%% Title     %%%%%%%%%
%%%%%%%%%%%%%%%%%%%%%%%%%
\renewcommand{\thefootnote}{\fnsymbol{footnote}}
\setcounter{footnote}{1}

% %%%%%%% CHOOSE TITLE PAGE--------
%\onecolumn
%\input{title-LHCb-INT}
%\input{title-LHCb-ANA}
%\input{title-LHCb-CONF}
%\input{title-LHCb-FIGURE}
% ===============================================================================
% Purpose: LHCb-PAPER journal paper title page template
% Author: 
% Created on: 2010-09-25
% ===============================================================================

%%%%%%%%%%%%%%%%%%%%%%%%%
%%%%%  TITLE PAGE  %%%%%%
%%%%%%%%%%%%%%%%%%%%%%%%%
\begin{titlepage}
\pagenumbering{roman}

% Header ---------------------------------------------------
\vspace*{-1.5cm}
\centerline{\large EUROPEAN ORGANIZATION FOR NUCLEAR RESEARCH (CERN)}
\vspace*{1.5cm}
\noindent
\begin{tabular*}{\linewidth}{lc@{\extracolsep{\fill}}r@{\extracolsep{0pt}}}
\ifthenelse{\boolean{pdflatex}}% Logo format choice
{\vspace*{-1.5cm}\mbox{\!\!\!\includegraphics[width=.14\textwidth]{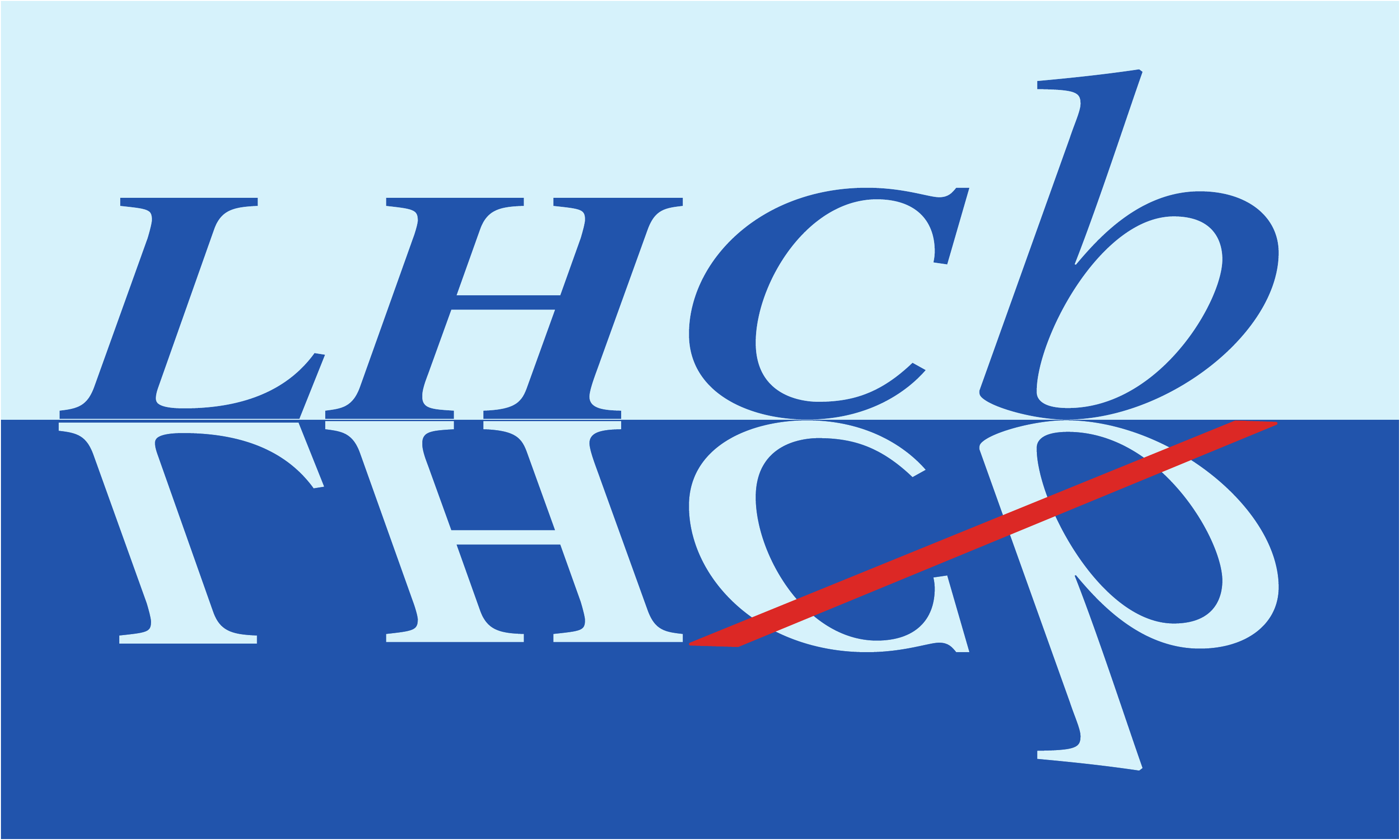}} & &}%
{\vspace*{-1.2cm}\mbox{\!\!\!\includegraphics[width=.12\textwidth]{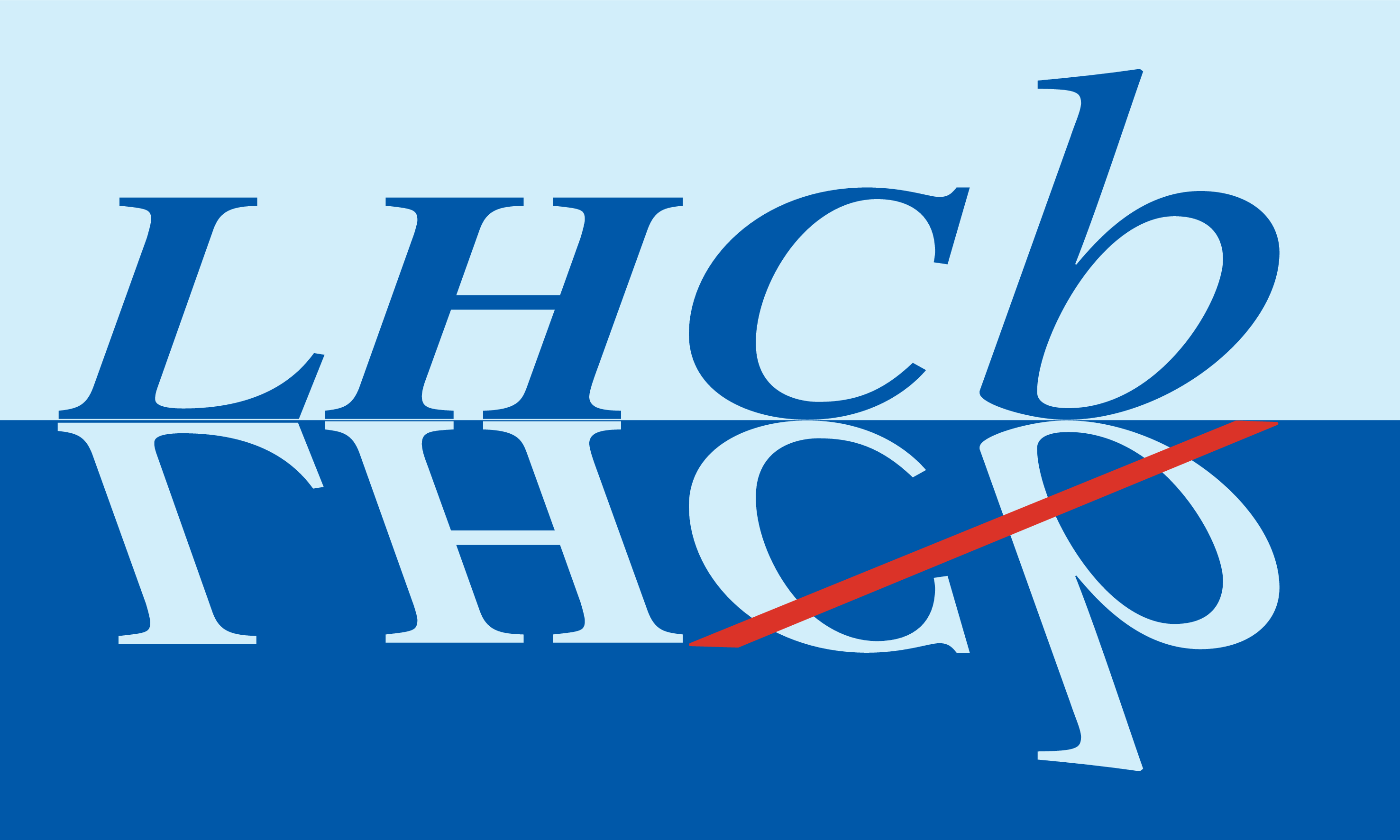}} & &}%
\\
 & & CERN-EP-2023-241 \\  % ID 
 & & LHCb-PAPER-2023-029 \\  % ID 
 & & February 16, 2024 \\ % Date - Can also hardwire e.g.: 23 March 2010
 & & \\
% not in paper \hline
\end{tabular*}

\vspace*{4.0cm}

% Title --------------------------------------------------
{\normalfont\bfseries\boldmath\huge
\begin{center}
% DO NOT EDIT HERE. Instead edit macro in main.tex to keep metadata correct
  \papertitle 
\end{center}
}

\vspace*{2.0cm}

% Authors -------------------------------------------------
\begin{center}
%In the footnote, replace 'paper' by 'Letter' in case of submission to PRL or PLB 
% Edit macro in main.tex to keep metadata correct
\paperauthors\footnote{Authors are listed at the end of this paper.}
\end{center}

\vspace{\fill}

% Abstract -----------------------------------------------
\begin{abstract}
  \noindent
A measurement of \CP-violating observables in \Bpm \to \Dstar \Kpm and \Bpm \to \Dstar \pipm decays is made where the photon or neutral pion from the \Dstar\to \D\g or \Dstar\to\D\piz decay is not reconstructed. The \D meson is reconstructed in the self-conjugate decay modes, ${\D \to \KS \pip \pim}$ or ${\D \to \KS \Kp \Km}$. The distribution of signal yields in the \D decay phase space is analysed in a model-independent way. The measurement uses a data sample collected in proton-proton collisions at centre-of-mass energies of 7, 8, and 13\tev, corresponding to a total integrated luminosity of approximately 9\invfb. The ${\Bpm \to \Dstar \Kpm}$ and ${\Bpm \to \Dstar \pipm}$ \CP-violating observables are interpreted in terms of hadronic parameters and the CKM angle \g, resulting in a measurement of $\g = (92^{+21}_{-17})\degrees$. The total uncertainty includes the statistical and systematic uncertainties, and the uncertainty due to external strong-phase inputs.
  
\end{abstract}

\vspace*{1.0cm}

\begin{center}
  Published in JHEP 02 (2024) 118

\end{center}

\vspace{\fill}

{\footnotesize 
% Edit macro in main.tex to keep metadata correct
\centerline{\copyright~\papercopyright. \href{\paperlicenceurl}{\paperlicence}.}}
\vspace*{2mm}

\end{titlepage}

%%%%%%%%%%%%%%%%%%%%%%%%%%%%%%%%
%%%%%  EOD OF TITLE PAGE  %%%%%%
%%%%%%%%%%%%%%%%%%%%%%%%%%%%%%%%

%  empty page follows the title page ----
\newpage
\setcounter{page}{2}
\mbox{~}
%\newpage
%
%% Author List ----------------------------
%%  You need to get a new author list!
%\input{Authorship_LHCb-PAPER-2023-029}
%
%The author list for journal publications is provided by the Membership Committee shortly after 'approval to go to paper' has been given.
%%It will be made available on the page
%%\verb!http://www.physik.uzh.ch/~strauman/forMemCo/LHCb-PAPER-XXXX-XXX/! .
%It will be sent to you by email shortly after a paper number has beens assigned.
%The author list should be included already at first circulation, 
%to allow new members of the collaboration to verify whether they have been included correctly.
%Occasionally a misspelled name is corrected or associated institutions become full members.
%In that case, a new author list will be sent to you.
%In case line numbering doesn't work well after including the authorlist, try moving the \verb!\bigskip! after the last author to a separate line.
%
%
%The authorship for Conference Reports should be ``The LHCb
%  collaboration'', with a footnote giving the name(s) of the contact
%  author(s), but without the full list of collaboration names.

%\twocolumn
% %%%%%%%%%%%%% ---------

\renewcommand{\thefootnote}{\arabic{footnote}}
\setcounter{footnote}{0}

%%%%%%%%%%%%%%%%%%%%%%%%%%%%%%%%
%%%%%  Table of Content   %%%%%%
%%%%%%%%%%%%%%%%%%%%%%%%%%%%%%%%
%%%% Uncomment if desired
%\tableofcontents
\cleardoublepage

%%%%%%%%%%%%%%%%%%%%%%%%%
%%%%% Main text %%%%%%%%%
%%%%%%%%%%%%%%%%%%%%%%%%%

\pagestyle{plain} % restore page numbers for the main text
\setcounter{page}{1}
\pagenumbering{arabic}

%% Uncomment during review phase. 
%% Comment before a final submission.
%\linenumbers

%% This is the main body
%% It is useful to have a single file so comemnts are not missed in overleaf.
\section{Introduction}
\label{sec:Introduction}

In the Standard Model, \CP violation in the quark sector is described using the Cabibbo-Kobayashi-Maskawa (CKM) matrix~\cite{Cabibbo:1963yz, Kobayashi:1973fv}. One representation of this matrix is the Unitarity Triangle~\cite{wolfenstein1983parametrization}. The CKM angle $\gamma \equiv$ arg$(-\Vud \Vub^{*}/\Vcd \Vcb^{*})$ can be measured using tree-level decays. Assuming tree-level processes do not include physics beyond the Standard Model~\cite{Brod_2015,Lenz:2019lvd}, a comparison of direct \g measurements and indirect determinations can test the Standard Model~\cite{blanke2019emerging}. This is because indirect determinations come from global CKM fits using other CKM observables, which are sensitive to physics beyond the Standard Model. In addition, measurements of \g have negligible theoretical uncertainty~\cite{Brod_2014} because all hadronic parameters are determined directly from data, removing the need to model non-perturbative hadronic effects that can result in large theoretical uncertainties. The current value of the indirect determination is ${\g=(65.5^{+1.1}_{-2.7})\degrees}$~\cite{CKMfitter2005} whilst the latest \g combination using \lhcb data yields $\g=(63.8^{+3.5}_{-3.7})\degrees$~\cite{LHCb-CONF-2022-003}.

The interference of \bquark \to ~\cquark and \bquark \to ~\uquark amplitudes can be used to measure \g. This paper presents a study of $\Bpm \to [\D\g/\piz]_{\Dstar} h^{\pm}$ decays, $\hadron = \{\pi, \kaon\}$. The \Dstar meson is a superposition of the $\Dstarz(2007)^{0}$ and $\Dstarzb(2007)^{0}$ states and decays to a neutral \D meson and an accompanying photon or neutral pion. The \D meson is a superposition of the \Dz and \Dzb states and subsequently decays to $\KS h^{+}h^{-}$. Therefore, the effects of the interference manifest themselves as differences in the \D-decay Dalitz-plot distributions for decays originating from a \Bp or \Bm meson. The $\D \to \KS h^+ h^-$ modes are considered golden channels given their rich interference structures. These can be exploited to measure \g~\cite{bondar2002proceedings, Bondar_2006, Bondar_2008, Giri_2003} by binning the Dalitz plots. To interpret the Dalitz-plot distributions in terms of \g, strong-phase information for the \D decays is required. In this study, a model-independent approach is adopted, which uses strong-phase measurements from \besiii~\cite{Ablikim_2020pi, Ablikim_2020K} and \cleo~\cite{Libby_2010}. The strong-phase parameters were determined using quantum correlated \Dz\Dzb pairs produced at the \psiprpr resonance. The use of these directly measured parameters results in systematic uncertainties on \g that are reliably determined. This avoids use of any amplitude model where the associated uncertainty is difficult to estimate.

The decay chain of $\Bpm\to\Dstar\Kpm$ with $\Dstar \to \D\piz$ or $\Dstar \to \D \gamma$, followed by the ${\D \to \KS h^+ h^-}$ decay, has been studied in the \belle~\cite{PhysRevD.81.112002} and \babar~\cite{PhysRevLett.105.121801} experiments. For each experiment, it was analysed alongside other $\Bpm$ decays using a model-dependent method. Recently, this decay has also been studied at LHCb~\cite{LHCb-PAPER-2023-012}, leading to a measurement of $\gamma=(69\pm 14)\degrees$. The key difference between the measurement presented in Ref.~\cite{LHCb-PAPER-2023-012} and that presented here is the treatment of the photon and neutral pion from the decay of the \Dstar meson. Here, no reconstruction requirements are placed on these neutral particles, leading to a high selection efficiency, but with the additional complication of a higher background level. The $\Bpm \to \Dstar\Kpm$ \CP-violating observables are better constrained here, but the large background contributions lead to, comparatively, worse constraints on the $\Bpm \to\Dstar\pipm$ \CP-violating observables. In Ref.~\cite{LHCb-PAPER-2023-012}, the photon and neutral pion from the \Dstar decay are reconstructed and the selection requirements placed on them reduce the signal efficiency by approximately 75\% due to the complex reconstruction of low momentum neutral particles at \lhcb~\cite{LHCbDetector2015}. Since no requirement is placed on the neutral particle in the present analysis, the $\Dstar \to \D\g$ and $\Dstar \to \D \piz$ components are distinguished using their different distributions in the invariant mass combination of the \D meson and its companion hadron, $m(\D\hadron)$. This exploits the spin-parity structure of the \Dstar decays. The measurement in this analysis is performed using a data sample collected in proton-proton ($pp$) collisions at centre-of-mass energies of 7, 8, and 13 \tev, corresponding to a total integrated luminosity of approximately 9\invfb.

\section{Analysis overview}
\label{sec:AnalysisOverview}
In $\Bm \to [\D\g/\piz]_{\Dstar} h^{-}$ decays, the \D meson is a superposition of the \Dz and \Dzb states. Therefore, the favoured contribution (via \Dz) and the suppressed contribution (via \Dzb) can interfere if the decay products of the \D meson are common to both \Dz and \Dzb. As an example, the overall amplitude for $\Bm \to [\D\piz]_{\Dstar} \Km$ can be written as,
\begin{equation}
        \mathcal{A}(\Bm \to [\D\piz]_{\Dstar} \Km) = \mathcal{A}_{\B}(\mathcal{A}_{\D} + r_{\B}^{\Dstar\kaon}\exp[i(\delta_{\B}^{\Dstar\kaon} - \g)]\mathcal{A}_{\Db}),
        \label{eq:interference_dpi0}
\end{equation}
and for $\Bm \to [\D\g]_{\Dstar} \Km$ as
\begin{equation}
        \mathcal{A}(\Bm \to [\D\g]_{\Dstar} \Km) = \mathcal{A}_{\B}(\mathcal{A}_{\D} -r_{\B}^{\Dstar\kaon}\exp[i(\delta_{\B}^{\Dstar\kaon} - \g)]\mathcal{A}_{\Db}),
        \label{eq:interference_dgam}
\end{equation}
where $\mathcal{A}_{\B}$ is the amplitude of the favoured \Bm decay, and $\mathcal{A}_{\D(\Db)}$ is the amplitude of the \Dz(\Dzb) decay. The hadronic parameters, $r_{\B}^{\Dstar\kaon}$ and $\delta_{\B}^{\Dstar\kaon}$, are the ratio of the magnitudes of the suppressed and favoured \Bpm decays and the strong-phase difference between them, respectively. The sign change between Eqs.~\ref{eq:interference_dpi0} and \ref{eq:interference_dgam} arises from a phase shift of $\pi$ between the strong-phase differences due to the quantum numbers of the pion and photon~\cite{Bondar_2004}. In order to use a single strong-phase parameter for all $\Bpm \to [\D \g/\piz]_{\Dstar} \Kpm$ decays, the phase shift is propagated through as a sign change. Similar amplitudes to Eqs.~\ref{eq:interference_dpi0} and~\ref{eq:interference_dgam} can be written for \Bp decays where $\mathcal{A}_{\D}$ is replaced by $\mathcal{A}_{\Db}$ and vice versa, and $-\g$ is replaced by $+\g$. This sign change on \g allows a measurement to be extracted using \Bp and \Bm decays. Similar amplitudes can also be written for the \Bpm \to \Dstar \pipm with replacements for $r_{B}$ and $\delta_{B}$.

The Dalitz-plot formalism is used to describe the $\D \to \KS \hadron^{+}\hadron^{-}$ phase space with two degrees of freedom: $s_{+}$, the squared invariant mass of the $\KS \hadron^{+}$ system, and $s_{-}$, the squared invariant mass of the $\KS \hadron^{-}$ system. Each point on the Dalitz plot represents a different kinematic final state of the \D decay described by the corresponding hadronic parameters. The analysis is performed in $2 \times\mathcal{N}$ bins of the Dalitz plot (referred to as Dalitz bins), labelled $i = -\mathcal{N}$ to $+\mathcal{N}$ excluding zero. The Dalitz bins are symmetrical about the line $s_{+} = s_{-}$, with positive bins where $s_{-} > s_{+}$ and negative bins where $s_{-} < s_{+}$. The Dalitz binning schemes for $\D \to \KS  \pip \pim$ and $\D \to \KS  \Kp \Km$ decays are known as the `optimal' ($\mathcal{N}=8$) and `2-bins' ($\mathcal{N}=2$) binning schemes, respectively. These were constructed such that the average of the strong phases across a bin maximises sensitivity to \g and are detailed in Ref.~\cite{Libby_2010}. These are the same binning schemes used in Ref.~\cite{LHCb-PAPER-2020-019} which studied $\Bpm \to \D\hadron^{\pm}$ decays with the \D meson decaying to the same final state.

The analysis presented in this paper uses a model-independent approach, taking strong-phase measurements for the \D decays from \besiii~\cite{Ablikim_2020pi, Ablikim_2020K} and \cleo~\cite{Libby_2010} in the form of $c_{i}$ and $s_{i}$ parameters. These are the amplitude-averaged cosine and sine of the strong-phase difference between the \Dz and \Dzb decays,
\begin{align}
\begin{split}
c_{i} \equiv \frac{\int_{i} \mathrm{d}s_{-}\mathrm{d}s_{+} \left|\mathcal{A}_{\D}(s_{-}, s_{+})\right| \left| \mathcal{A}_{\Db}(s_{+}, s_{-}) \right|\cos(\delta_{\D}(s_{-}, s_{+}) - \delta_{\D}(s_{+}, s_{-}))}{\sqrt{\int_{i} \mathrm{d}s_{-}\mathrm{d}s_{+}\left|\mathcal{A}_{\D}(s_{-}, s_{+})\right|^{2} \int_{i} \mathrm{d}s_{-}\mathrm{d}s_{+}\left| \mathcal{A}_{\Db}(s_{+}, s_{-}) \right|^{2}}}, \\
s_{i} \equiv \frac{\int_{i} \mathrm{d}s_{-}\mathrm{d}s_{+} \left|\mathcal{A}_{\D}(s_{-}, s_{+})\right| \left| \mathcal{A}_{\Db}(s_{+}, s_{-}) \right|\sin(\delta_{\D}(s_{-}, s_{+}) - \delta_{\D}(s_{+}, s_{-}))}{\sqrt{\int_{i} \mathrm{d}s_{-}\mathrm{d}s_{+}\left|\mathcal{A}_{\D}(s_{-}, s_{+})\right|^{2} \int_{i} \mathrm{d}s_{-}\mathrm{d}s_{+}\left| \mathcal{A}_{\Db}(s_{+}, s_{-}) \right|^{2}}},
\end{split}
\end{align}
where $\int_{i} \mathrm{d}s_{-}\mathrm{d}s_{+}$ denotes integration over the phase space of the $i^{\text{th}}$ Dalitz bin. These measurements are efficiency-corrected at \besiii and \cleo and assumed to be directly applicable in an \lhcb analysis as the efficiency correction is small.

In this analysis, effects from \CP violation in \D decays and from \D mixing are assumed to be negligible, considering the current experimental sensitivity~\cite{PhysRevD.82.034033}. In addition, \KS mixing, regeneration in \KS interactions with matter, and \KS~\CP violation are similarly ignored~\cite{Bj_rn_2019}. As a result, the \CP transformation of the \D decay amplitudes results in $\mathcal{A}_{\D}(s_{-}, s_{+}) = \mathcal{A}_{\Db}(s_{+}, s_{-})$. The expected yield of the \Bpm decays per Dalitz bin is calculated by squaring the overall amplitudes and integrating over phase space and decay time. When doing so, the observed fractional yield of \Dz decays in bin $i$ is defined as,
\begin{equation}
F_{i} \equiv \frac{\int_{i} \mathrm{d}s_{-}\mathrm{d}s_{+} \eta(s_{-}, s_{+}) \left|\mathcal{A}_{\D}(s_{-}, s_{+})\right|^{2}}{\int \mathrm{d}s_{-}\mathrm{d}s_{+}\eta(s_{-}, s_{+})\left|\mathcal{A}_{D}(s_{-}, s_{+})\right|^{2}},
\end{equation}
where $\eta(s_{-}, s_{+})$ is the experimental efficiency. Furthermore, a set of \CP-violating observables can be defined using the physical parameters, $r_{\B}^{\Dstar \kaon}$, $\delta_{\B}^{\Dstar \kaon}$, and \g,
\begin{equation}
x^{\Dstar\kaon}_{\pm} \equiv r_{\B}^{\Dstar\kaon}\cos(\delta_{\B}^{\Dstar\kaon} \pm \g) {\rm\ \ and\ } \; 
y^{\Dstar\kaon}_{\pm} \equiv r_{\B}^{\Dstar\kaon}\sin(\delta_{\B}^{\Dstar\kaon} \pm \g).
\label{eq:CPobs_def}
\end{equation}
A parameterisation~\cite{ PhysRevD.102.053003} for the $\Bpm \to \Dstar \pipm$ decays, because \g is common among all decays, involves the introduction of a new quantity,
\begin{equation}
    \xi \equiv \frac{r_{\B}^{\Dstar\pi}}{r_{\B}^{\Dstar\kaon}} \exp[i(\delta_{\B}^{\Dstar\pi} - \delta_{\B}^{\Dstar\kaon})].
    \label{eq:csi}
\end{equation}
Using the real and imaginary parts of the $\xi$ parameter, the \CP-violating observables for $\Bpm \to \Dstar \pipm$ can be recovered,
\begin{equation}
x_{\pm}^{\Dstar\pi} = {\Re(\xi)}x_{\pm}^{\Dstar\kaon} -  {\Im(\xi)}y_{\pm}^{\Dstar\kaon} {\rm\ \ and\ } \;
y_{\pm}^{\Dstar\pi} = {\Re(\xi)}y_{\pm}^{\Dstar\kaon} + {\Im(\xi)}x_{\pm}^{\Dstar\kaon}.
  \end{equation}
The resulting yield equations for $\Bpm \to [\D\piz/\g]_{\Dstar} \hadron^{\pm}$ decays with the appropriate \CP-violating observables, where the superscripts have been omitted for brevity, are
\begin{align}
    \begin{split}
        N^{+\text{, }\D\piz}_{i} &= h_{\B}^{+\text{, }\D\piz}[F_{- i} + (x_{+}^{2} + y_{+}^{2})F_{+ i} + 2\sqrt{F_{- i}F_{+ i}}(c_{i}x_{+} - s_{i}y_{+})],\\
        N^{-\text{, }\D\piz}_{i} &= h_{\B}^{-\text{, }\D\piz}[F_{+ i} + (x_{-}^{2} + y_{-}^{2})F_{- i} + 2\sqrt{F_{+ i}F_{- i}}(c_{i}x_{-} + s_{i}y_{-})],\\
        N^{+\text{, }\D\g}_{i} &= h_{\B}^{+\text{, }\D\g}[F_{- i} + (x_{+}^{2} + y_{+}^{2})F_{+ i} - 2\sqrt{F_{- i}F_{+ i}}(c_{i}x_{+} - s_{i}y_{+})],\\
        N^{-\text{, }\D\g}_{i} &= h_{\B}^{-\text{, }\D\g}[F_{+ i} + (x_{-}^{2} + y_{-}^{2})F_{- i} - 2\sqrt{F_{+ i}F_{- i}}(c_{i}x_{-} + s_{i}y_{-})],
    \end{split}
    \label{eq:SigYieldEqns}
\end{align}
where $h_{\B}^{\pm\text{, }\D\piz}$ and $h_{\B}^{\pm\text{, }\D\g}$ are normalisation constants. The separate normalisation for each \B charge and $\Dstar$ decay absorbs production and most detection asymmetries. To fix this normalisation, the yield per Dalitz bin is determined using,
\begin{equation}
 N^{\pm}_{ i} = \frac{Y^{\pm}_{i}}{\sum_{j}Y^{\pm}_{j}} N^{\pm}_{\text{total}},
 \label{eq:Yi}
\end{equation}
where $Y^{\pm}_{i}$ is the fractional yield in a Dalitz bin, \ie Eq.~\ref{eq:SigYieldEqns} without the normalisation constants, and is therefore determined via the $F_{i}$ parameters and \CP-violating observables. The total yield for a signal component over the entire \D decay phase space is determined in a fit to the reconstructed mass and denoted by $N^{\pm}_{\text{total}}$. In deriving Eq.~\ref{eq:SigYieldEqns}, the effect of detector efficiency is taken into account in the measured $F_i$ values.

From the above description, the analysis strategy can be summarised as follows. A sample is selected as described in Sec.~\ref{sec:Selection}. Then the $m(\D\hadron)$ distribution of every signal and background component is determined using a global fit, as described in Sec.~\ref{sec:GlobalFit}. The resulting mass shapes and yields are used in a subsequent fit to determine the \CP-violating observables, as described in Sec.~\ref{sec:CPFit}. The associated systematic uncertainties are evaluated in Sec.~\ref{sec:Systematics}. The \CP-violating observables are then interpreted in terms of the physical parameters, including \g, as described in Sec.~\ref{sec:Interpretation}.

\section{\lhcb detector and simulation}
\label{sec:Detector}
The \lhcb detector~\cite{LHCb-DP-2008-001,LHCb-DP-2014-002} is a single-arm forward
spectrometer covering the \mbox{pseudorapidity} range $2<\eta <5$,
designed for the study of particles containing \bquark or \cquark
quarks. The detector includes a high-precision tracking system
consisting of a silicon-strip vertex detector surrounding the $pp$
interaction region, a large-area silicon-strip detector located
upstream of a dipole magnet with a bending power of about
$4{\mathrm{\,Tm}}$, and three stations of silicon-strip detectors and straw
drift tubes placed downstream of the magnet.
The tracking system provides a measurement of the momentum, \ptot, of charged particles with
a relative uncertainty that varies from 0.5\% at low momentum to 1.0\% at 200\gevc for tracks that are reconstructed in the vertex detector.
For these, the minimum distance of a track to a primary $pp$ collision vertex (PV), the impact parameter (IP), 
is measured with a resolution of $(15+29/\pt)\mum$,
where \pt is the component of the momentum transverse to the beam, in\,\gevc.
Different types of charged hadrons are distinguished using information
from two ring-imaging Cherenkov detectors. 
Photons, electrons and hadrons are identified by a calorimeter system consisting of
scintillating-pad and preshower detectors, an electromagnetic calorimeter
and a hadronic calorimeter. Muons are identified by a
system composed of alternating layers of iron and multiwire
proportional chambers. 

The online event selection is performed by a trigger,
which consists of a hardware stage, based on information from the calorimeter and muon
systems, followed by a software stage, which applies a full event
reconstruction. The latter is further split into two stages. First, at least one particle is required to have high \pt and high \chisqip, where \chisqip is defined as the difference in the primary vertex fit \chisq with and without the inclusion of that particle. 
Then a multivariate algorithm~\cite{BBDT} is used to identify secondary vertices consistent with being a two-, three-, or four-track $\bquark$-hadron decay. 
The PVs are fitted with and without the tracks of the decay products of the \B candidate, and the PV that gives the smallest \chisqip is associated with the \B candidate.

Simulation is primarily used to model mass distributions of the signal and background components and determine efficiencies. In the simulation, $pp$ collisions are generated using \pythia~\cite{Sjostrand:2007gs,*Sjostrand:2006za} with a specific \lhcb configuration~\cite{LHCb-PROC-2010-056}. Decays of unstable particles are described by \evtgen~\cite{Lange:2001uf}, in which final-state radiation is generated using \photos~\cite{davidson2015photos}. The interaction of the generated particles with the detector and its response are implemented using the \geant toolkit~\cite{Allison:2006ve, *Agostinelli:2002hh} as described in Ref.~\cite{LHCb-PROC-2011-006}. The underlying $pp$ interaction is reused multiple times, with an independently generated signal decay for each~\cite{LHCb-DP-2018-004}. In addition, fast simulation~\cite{Cowan:2016tnm,Back:2017zqt} is used to model backgrounds with broad mass distributions.

\section{Candidate selection}
\label{sec:Selection}
The candidate selection used in this paper is built on that used to select ${\Bpm\to[\KS h^{+}h^{-}]_{\D}h'^{\pm}}$ decays in Ref.~\cite{LHCb-PAPER-2020-019}. This reconstructs the same final state as in the present analysis, where reconstruction of the photon or neutral pion from the \Dstar decay is not required.

A $B$ decay candidate is built by first reconstructing a \KS meson using two tracks identified as oppositely charged pions. If the pion tracks are reconstructed in the vertex detector, they form a \textit{long} \KS candidate, otherwise they form a \textit{downstream} \KS candidate. This distinction is important since long \KS candidates have better mass, vertex, and momentum resolution. For this reason, long and downstream \KS candidates are treated separately. The \KS meson is combined with two further tracks identified as either oppositely charged pions or kaons to form a \D meson candidate. Particle identification (PID) requirements are applied to these charged kaons and pions. The \D meson candidate is then combined with a companion kaon or pion, again distinguished using PID requirements, to form a \B meson candidate. To improve the mass resolution of the reconstructed \B candidate, a kinematic fit~\cite{Hulsbergen:2005pu} is applied, where the reconstructed masses of the \KS and \D mesons are constrained to their known values.

The optimal PID requirement for the companion track is determined using pseudoexperiments to evaluate the sensitivity to \g. A tighter identification requirement for the $DK$ sample than that in Ref.~\cite{LHCb-PAPER-2020-019} is found to be favourable, resulting in around 70\% of true companion kaons reconstructed in the $\Bpm \to \D \Kpm$ data sample, with the remainder misidentified and reconstructed in the $\Bpm \to \D \pipm$ sample. The requirement results in only 5\% of the true companion pions being misidentified and reconstructed in the $\Bpm \to \D \Kpm$ data sample. Further PID requirements are placed on the \D decay products to reduce background from semileptonic \D decays.

The background contributions from $\Bpm \to \Dstar \hadron^{\pm}$ decays with ${\D \to \pip\pim\pip\pim}$ or ${\D \to \Kp\Km\pip\pim}$ are reduced by requiring the decay vertex of long \KS candidates to be significantly displaced from the \D decay vertex. Also, the background from \B decays that reach the final state without an intermediate \D meson is reduced by requiring well separated \D and \B decay vertices.

The large combinatorial background is suppressed using a boosted decision tree (BDT)~\cite{Breiman, AdaBoost} classifier. The details of this can be found in Ref.~\cite{LHCb-PAPER-2018-017}, with separate BDT classifiers for long and downstream \KS candidates and for data collected from 2011--2012 and 2015--2018. The optimal thresholds that maximise the sensitivity to \g are determined using pseudoexperiments. Compared to Ref.~\cite{LHCb-PAPER-2020-019}, which used the same BDT classifier, a tighter threshold is found to be necessary, due to the higher rate of combinatorial background in the region of the $\B\to \Dstar \hadron$ decays compared to $\B \to \D\hadron$ decays in the $m(\D\hadron)$ spectra. 

Reconstructed candidates are assigned to one of 8 categories according to the \B decay, \D decay, and \KS candidate type: $(\D\pi, \D \kaon) \times (\KS\pi\pi, \KS \kaon\kaon) \times (\text{downstream}, \text{long})$. A global fit is performed simultaneously in these categories. 

\section{Reconstructed mass fit}
\label{sec:GlobalFit}
The global fit is a binned extended maximum-likelihood fit to the reconstructed mass, $m(\D\hadron)$, over the range 4900--5600~\mevcc. There are four signal components, ${\Bpm \to [\D\g]_{\Dstar}\Kpm}$, ${\Bpm \to [\D\g]_{\Dstar}\pipm}$, ${\Bpm \to [\D\piz]_{\Dstar}\Kpm}$, and ${\Bpm \to [\D\piz]_{\Dstar}\pipm}$ where a photon or a neutral pion are missed, and a number of background components from both charged and neutral \B hadrons.

The signals from the two $\Dstar$ decay modes are distinguished using their different $m(\D\hadron)$ spectra which depend on the spin-parity and mass of the missing neutral particle. For $\Bpm \to [\D\g]_{\Dstar}h^\pm$ decays, the $m(\D\hadron)$ distribution is described by a parabola exhibiting a maximum, where the range is defined by kinematic limits, $a$ and $b$. Mass-dependent reconstruction and selection efficiencies require the introduction of a linear dependence defined by the asymmetry parameter, $\zeta$. Detector resolution effects are accounted for by convolving the parabola and asymmetry term with a resolution function, $p_{\text{Res}}(m|\vec{\theta})$, defined by a set of variables $\vec{\theta}$. This results in a broad structure described by the following probability density function (PDF),
\begin{equation}
    \Bpm \to [\D\g]_{\Dstar}h^\pm: p(m) = -\int^{b}_{a} \mathrm{d}\mu ( \mu - a)(\mu -b) p_{\text{Res}}(m|\vec{\theta}) \left(\frac{1 - \zeta}{b-a} \mu + \frac{b\zeta - a}{b - a}\right).
    \label{eq:DgamPDF}
\end{equation}
In the case of $\Bpm \to [\D\piz]_{\Dstar}\hadron^{\pm}$ decays, the $m(\D\hadron)$ distribution is described by a parabola that exhibits a minimum. Reconstruction and detector effects lead to a double-peaked structure with corresponding $a$, $b$, and $\zeta$ parameters taking into account the difference between $\Bpm \to [\D\g]_{\Dstar}h^\pm$ and $\Bpm \to [\D\piz]_{\Dstar}\hadron^{\pm}$ decays. It is described by the following PDF,
\begin{equation}
    \Bpm \to [\D\piz]_{\Dstar}h^\pm: p(m) = \int^{b}_{a} \mathrm{d} \mu \left( \mu - \frac{a+b}{2} \right)^{2} p_{\text{Res}}(m|\vec{\theta}) \left(\frac{1 - \zeta}{b-a} \mu + \frac{b\zeta - a}{b - a}\right).
    \label{eq:Dpi0PDF}
\end{equation}
The resolution function for these signals is that used for the fully reconstructed background decays, $\Bpm \to \D \hadron^{\pm}$. This is the sum of a Gaussian and a modified Gaussian function,
\begin{equation}
    p(m) = 
    \begin{cases}
    f\exp\left(\frac{-(\Delta m)^{2}(1 + \beta(\Delta m)^{2})}{2\sigma^{2} + \alpha_{L}(\Delta m)^{2}}\right) + (1 - f)p_{G}(m|\mu, \sigma), & \text{if $\Delta m<0$}\\
    f\exp\left(\frac{-(\Delta m)^{2}(1 + \beta(\Delta m)^{2})}{2\sigma^{2} + \alpha_{R}(\Delta m)^{2}}\right) + (1 - f)p_{G}(m|\mu, \sigma), & \text{if $\Delta m>0$},
    \end{cases}
    \label{eq:DhPDF}
\end{equation}
where $p_{G}(m|\mu, \sigma)$ is a Gaussian function and $\alpha_{L}$, $\alpha_{R}$, and $\beta$ are used to model radiative tails. The parameter $\Delta m$ is ${m - m_{\B}}$ where $m_{\B}$ is the $B$-meson mass, a freely varying parameter in the fit. For the signal modes and the $\Bpm \to \D \hadron^{\pm}$ background, the shape parameters are initially determined according to the full \lhcb simulation. To obtain a good-quality description of the simulated signal samples, the $\Bpm \to [\D\piz]_{\Dstar}\hadron^{\pm}$ signal is parameterised by summing two instances of the function described in Eq.~\ref{eq:Dpi0PDF}. In the fit to data some parameters are allowed to vary to account for resolution differences between simulation and data. These include the width parameters for the ${\Bpm \to [\D\piz]_{\Dstar}\hadron^{\pm}}$ signal and the fully reconstructed backgrounds, and the asymmetry parameters for the ${\Bpm \to [\D\g]_{\Dstar}h^{\pm}}$ signal. The asymmetry parameters for the $\Bpm \to [\D\piz]_{\Dstar}\hadron^{\pm}$ component are also allowed to vary. The remaining parameters are fixed from simulation.

The background contribution from $\Bz \to [\D \pimp]_{\Dstarmp}\hadron^{\pm}$ decays where the pion is missed has a similar decay topology to $\Bpm \to [\D\piz]_{\Dstar}\hadron^{\pm}$ decays. As a result, the background is modelled using the same parameterisation. In the fit to data the asymmetry and width parameters are shared with the $\Bpm \to [\D\piz]_{\Dstar}\hadron^{\pm}$ component, and other parameters are fixed from simulation.

There are a number of background contributions from various $\B$ decays, where one or more particles are not reconstructed and the companion hadron $h$ is correctly identified. The PDFs for these background components are determined from simulation. The functions with which the simulation samples are modelled are Eq.~\ref{eq:DgamPDF} or Eq.~\ref{eq:Dpi0PDF} with additional Gaussian components as appropriate, convolved with a resolution function. The resolution function is either a single or double Gaussian function as needed for a good description of simulated events. Simulated samples come from one of three sources: full simulation, which is typically used to generate the dominant two-body decays; a fast simulation~\cite{Cowan:2016tnm}, which is typically used to generate multi-body decays or less common two-body decays; and a second fast simulation package called \textsc{Laura}$^{++}$~\cite{Back:2017zqt}, which generates three-body decays from an amplitude model and then smears the generated momenta. The background components modelled by fast simulation have $m(\D\hadron)$ distributions which are typically very broad (on the order of 100\mevcc) compared to the $\Bpm \to \D\hadron^{\pm}$. This means that resolution differences between simulation and data are negligible and are not considered.

The dominant background contribution in $m(\D\pi)$ is from $B^{\pm} \to \D \pipm \piz$ decays, specifically due to $\Bpm \to \D \rhopm$ decays for which the PDF is determined from simulation. In $m(\D\kaon)$ the largest background is from $\B^{\pm} \to \D \Kpm \piz$ decays. This is modelled using fast simulation~\cite{Cowan:2016tnm} for the ${\Bpm \to \D [\Kpm \piz]_{K^*(892)^\pm}}$ and ${\Bpm \to \Kpm [\D\piz]_{\Dstar(2400)}}$ resonances. These resonances are combined into a single sample using fit fractions from Ref.~\cite{LHCb-PAPER-2015-059} with branching ratios from Ref.~\cite{PDG2022}. The $m(\D\hadron)$ distributions for ${\B \to \Dstar \hadron \pi}$ decays are determined from simulation for the ${\Bpm \to \Dstar \rhopm}$ and ${\Bpm \to \Dstar \Kstarpm}$ resonances, where the former is used to model contributions from both $\Bpm \to \Dstar \pipm \piz$ and $\Bz \to \Dstar \pip \pim$ decays.

Background contributions from $\Bs \to \D \kaon \pi$ and $\Bz \to \D \hadron \pi$ decays are generated using \textsc{Laura}$^{++}$ with amplitude models as determined in Refs.~\cite{LHCb-PAPER-2014-036, LHCb-PAPER-2015-059, LHCb-PAPER-2015-017}. A number of $m(\D\hadron)$ components are parameterised using fast simulation~\cite{Cowan:2016tnm}, namely, for $\Bs \to \Dstar \kaon \pi$ and nonresonant $\Bpm \to \D \pipm \piz$ decays, as well as $\Bpm \to \D a_{1}^{\pm}$ decays with $a_{1}^{\pm} \to \rhoz \pipm$. In the latter contribution, only the component where two pions with same-sign charges are missed is considered, since this leads to a background component that requires careful distinction in the subsequent \CP fit. The component where two oppositely charged pions are missed is similar in shape to the $\Bpm \to \Dstar \pipm \piz$ and combinatorial background components and is therefore absorbed by them. There is also a small background contribution from $\Lb \to \D p \pim$ decays, where the pion is missed and the proton is misidentified as a kaon. The $m(\D\hadron)$ distribution for this component is taken from Ref.~\cite{LHCb-PAPER-2020-036}. 

For each component, there is an associated cross-feed component where the companion hadron is misidentified. Simulated samples are modified to assign the wrong particle hypothesis and weighted to take into account the momentum-dependent misidentification efficiency. The resulting shape is then fitted with a number of Gaussian and/or Crystal Ball functions~\cite{Skwarnicki:1986xj} as required to achieve a high-quality fit. The combinatorial background is parameterised using an exponential function where the slope is allowed to vary freely in the fit to data.

For each component, a yield is specified in the fit. The yields for the fully reconstructed $\Bpm \to \D \pipm$ decays vary freely in each category. The signal yields and those corresponding to the $\Bpm \to \D \Kpm$ background decays are parameterised in terms of the $\Bpm \to \D \pipm$ yields with ratios of branching fractions and efficiencies. For the $\Bz \to [\D \pimp]_{\Dstarmp}\hadron^{\pm}$ background, the ratios are Gaussian-constrained to the values and uncertainties for the efficiencies, determined from simulation, and for branching fractions from Ref.~\cite{PDG2022}. For the majority of the remaining background components, their branching fractions and efficiencies are not known reliably and carry large uncertainties. A more accurate parameterisation uses fitted yields from Ref.~\cite{LHCb-PAPER-2020-036} where \CP violation in the global yields is assumed to be negligible. For example, this is a good assumption for the Cabibbo-favoured $\Bpm \to \D \pipm$ and $\Bs \to \Dzb \Km \pip$ decays where the corresponding value of $r_{B}$ is small. Therefore, the yield for the $\Bs \to \Dzb \Km \pip$ background can be parameterised as
\begin{equation}
N(\Bs \to \Dzb \Km \pip)_{\KS \hadron \hadron} = \frac{N(\Bs \to \Dzb \Km \pip)_{\pi \kaon}}{N(\Bm \to \Dz \pim)_{\kaon \pi}} \times N(\Bm \to \Dz \pim)_{\KS \hadron \hadron},
\label{eq:Bs_DKpi_yield_param}
\end{equation}
where charge-conjugate processes are implied. The \KS \hadron \hadron subscript denotes yields in this analysis, $\pi \kaon$ denotes yields from decays reconstructed as $\D(\to \pipm \Kmp)\Kpm$ in Ref.~\cite{LHCb-PAPER-2020-036}, and $\kaon \pi$ denotes yields from decays reconstructed as $\D(\to \Kpm \pimp)\pipm$ in Ref.~\cite{LHCb-PAPER-2020-036}. The same parameterisation is also used for the following backgrounds: ${\Bs \to \Dstar \Kmp \pipm}$, ${\Bz \to \D \pip \pim}$, ${\Bpm \to \D \hadron^{\pm} \piz}$, ${\B \to \Dstar \pi \pi}$, ${\Bpm \to \Dstar \Kpm \piz}$, ${\Bpm \to \D a_{1}^{\pm}}$, and ${\Lb \to \D p \pim}$.

The yield of the $\Bz \to \D^{(*)} \Kpm \pimp$ background is parameterised using that of the $\Bs \to \D^{(*)} \Kpm\pimp$ background using ratios of branching fractions, fragmentation fractions, and efficiencies due to the mass range. For the \Bpm \to ~\D\pipm \piz background, the total yield is parameterised similarly to Eq.~\ref{eq:Bs_DKpi_yield_param}, however the ratio of the \Bpm \to ~\D\rhopm resonance to the nonresonant component is unknown. Thus, the fraction between these two components varies freely in the fit to data and the resonant decay is found to be dominant. All cross-feed components are parameterised in terms of the yield of the correctly identified component and corrected for the misidentification rate. The yield of the combinatorial background is allowed to vary freely in each category.

\begin{figure}[tb]
    \centering
    \includegraphics[width=0.49\textwidth]{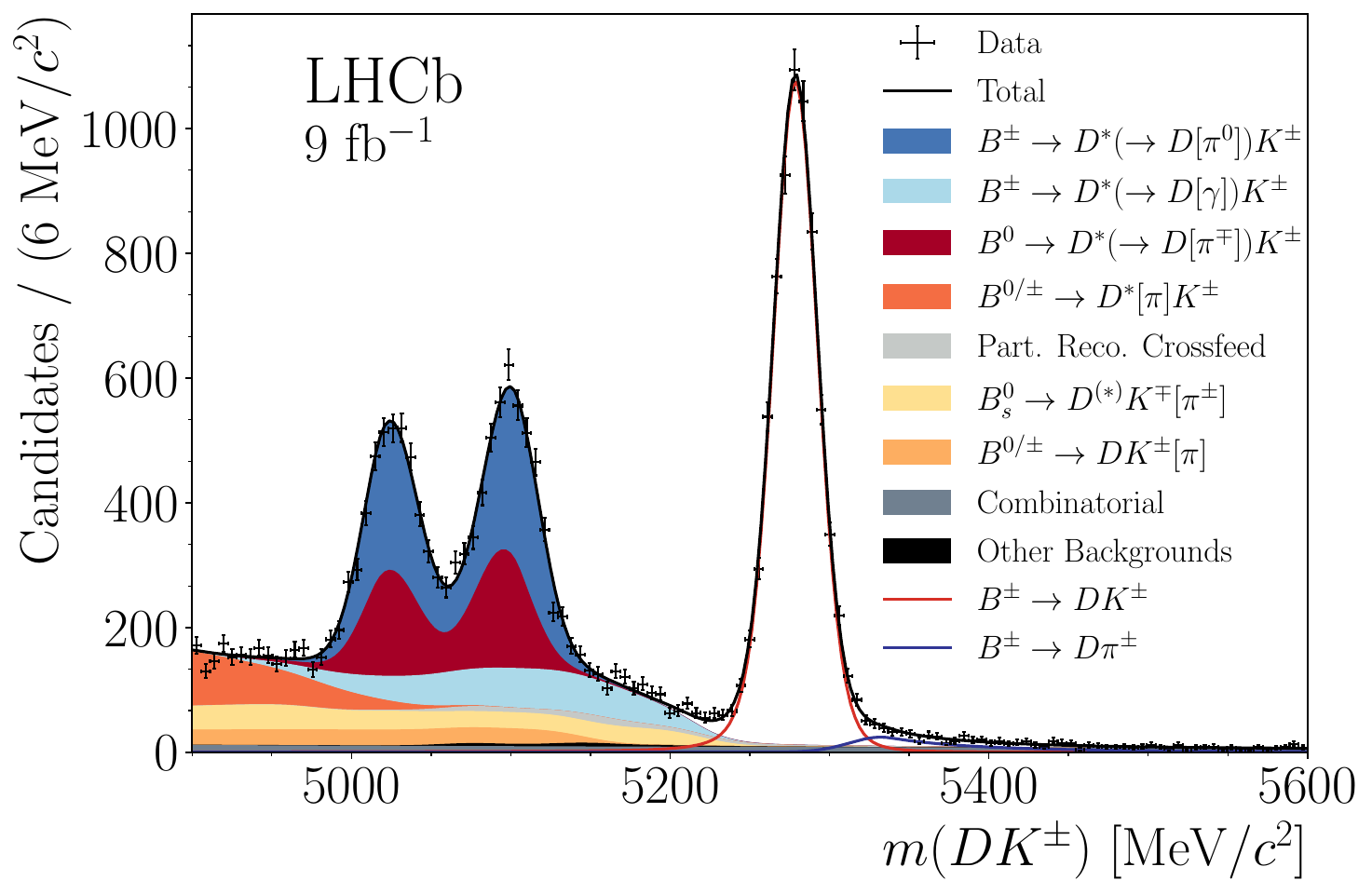} 
    \includegraphics[width=0.49\textwidth]{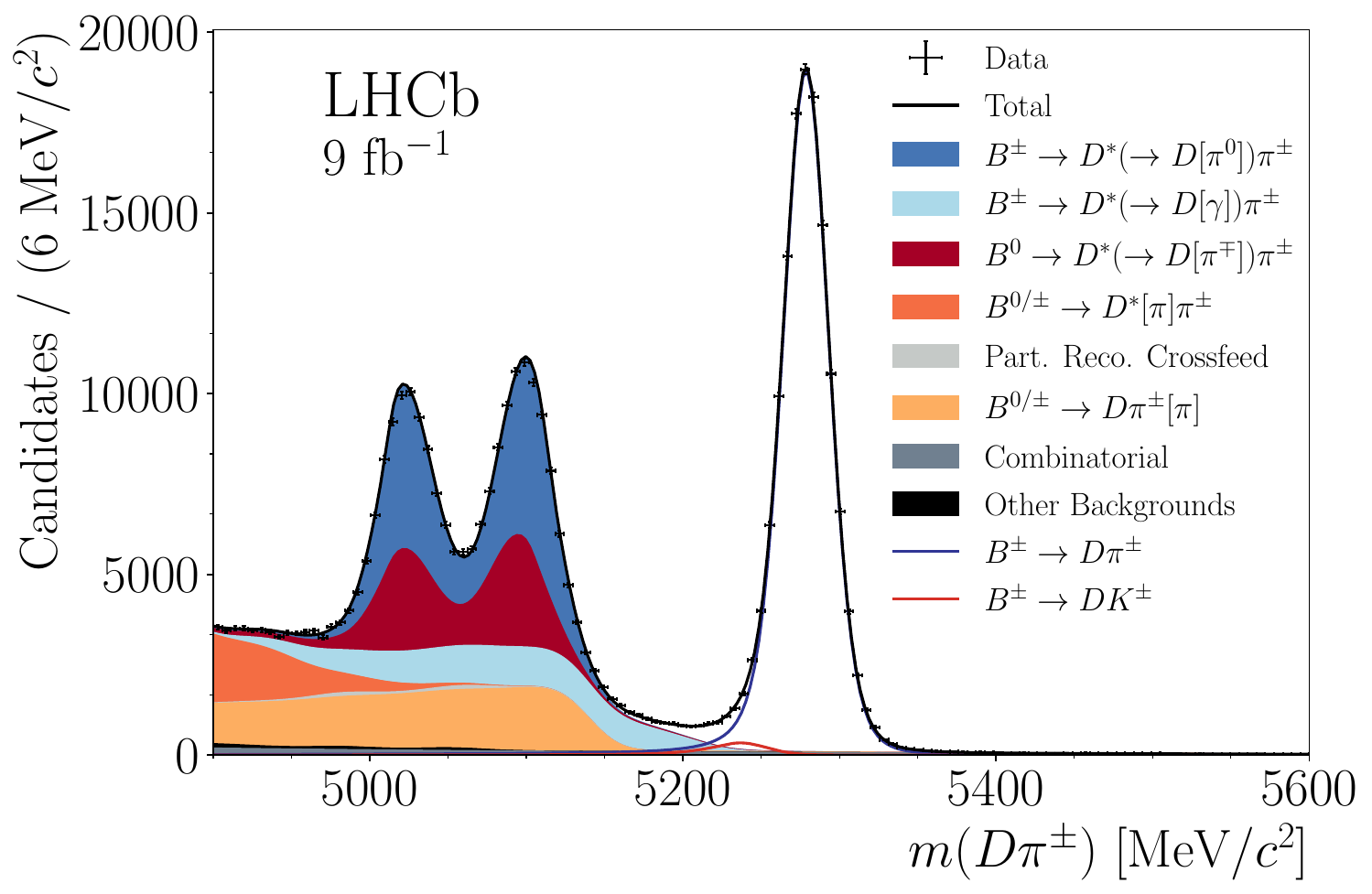}\\
    \includegraphics[width=0.49\textwidth]{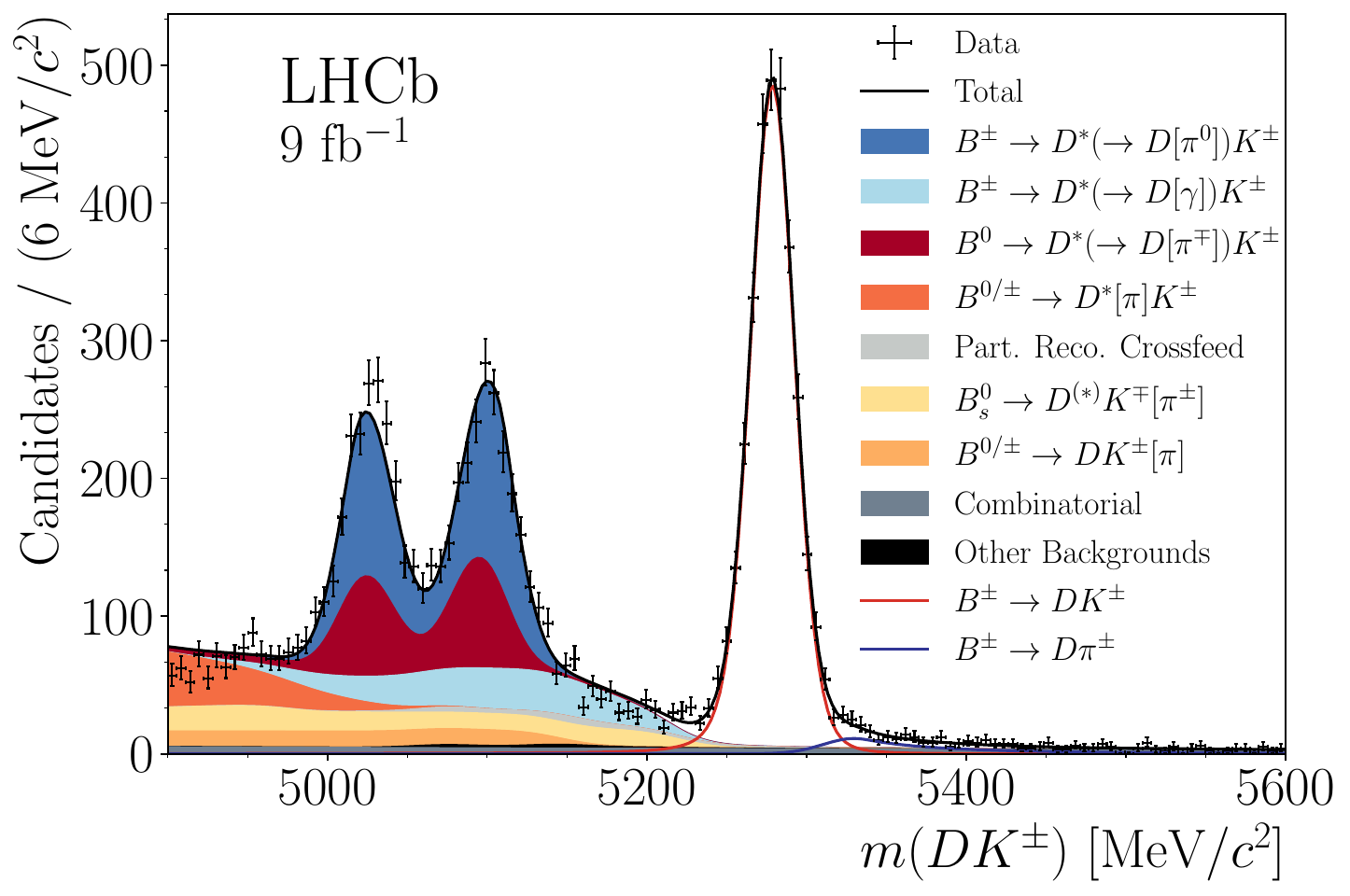} 
    \includegraphics[width=0.49\textwidth]{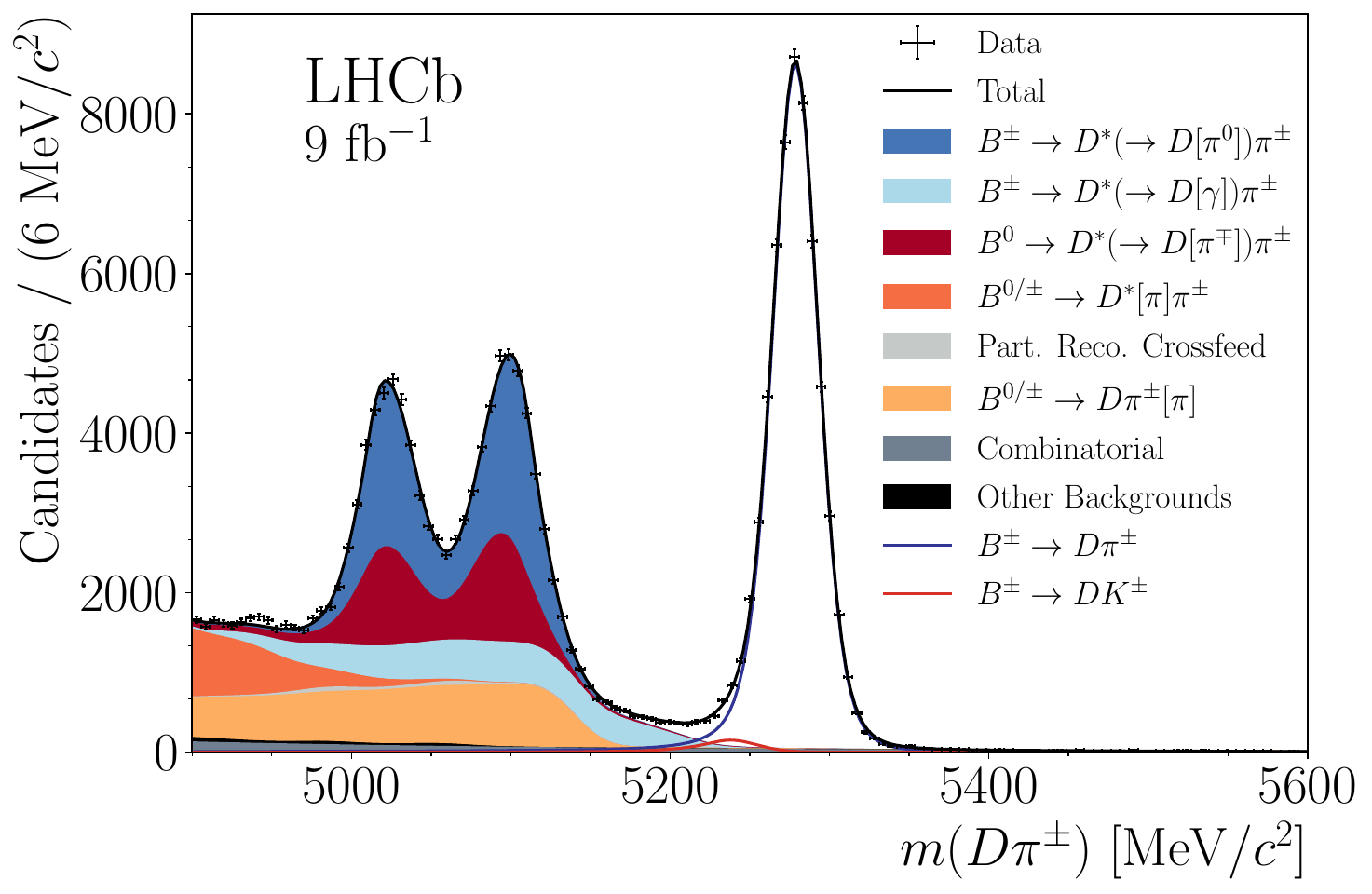}
    \caption{Mass distributions for (left) $\D\kaon$ samples and (right) $\D\pi$ samples, with $\D \to \KS \pip \pim$ decays reconstructed with a (top) downstream \KS candidate and a (bottom) long \KS candidate. The projections of the fit results are overlaid. In the legend, particles in square brackets are not reconstructed.}
    \label{fig:globalfit_results_kspp}
\end{figure}

\begin{figure}[tb]
    \centering
    \includegraphics[width=0.48\textwidth]{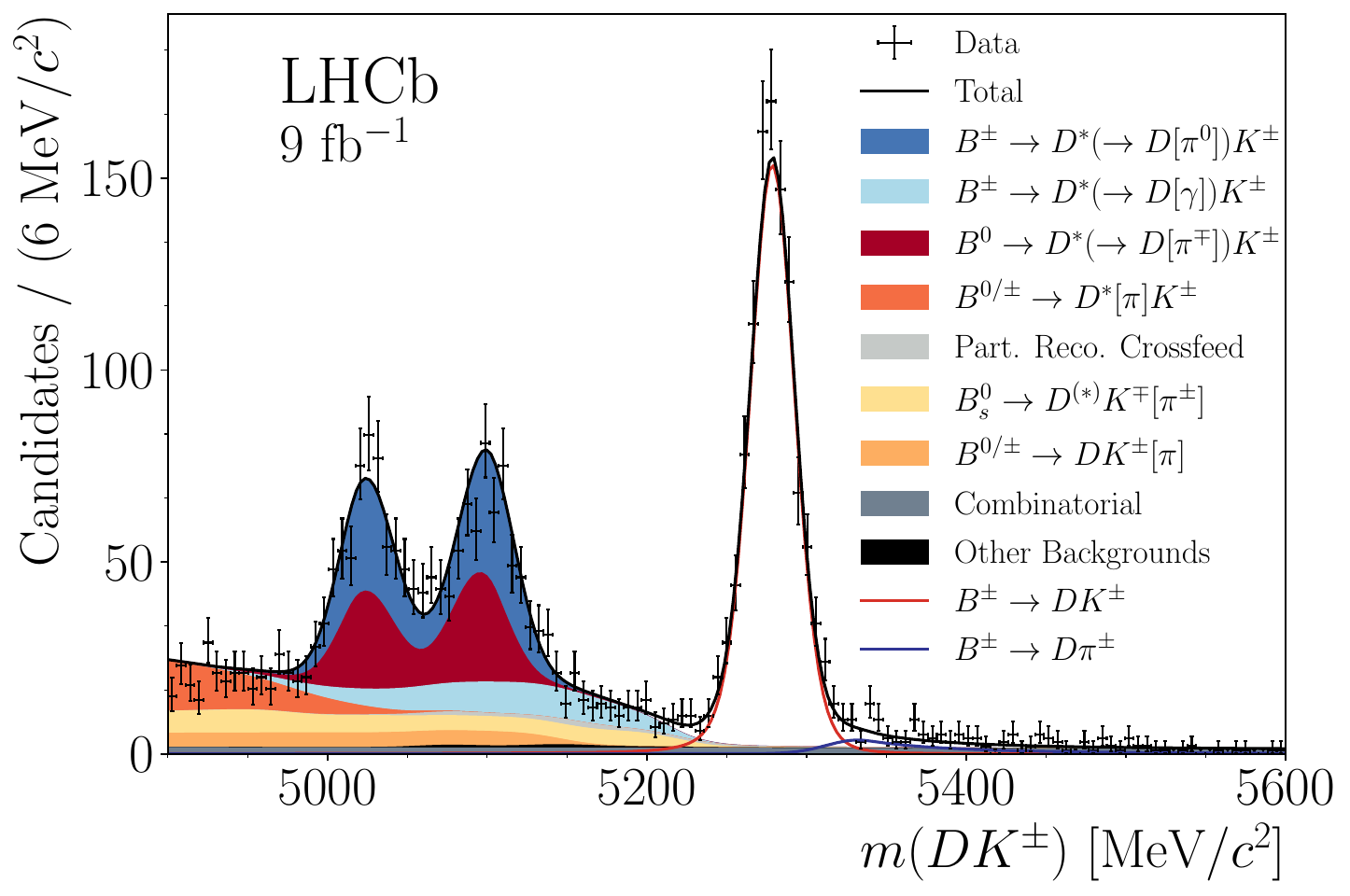} 
    \includegraphics[width=0.48\textwidth]{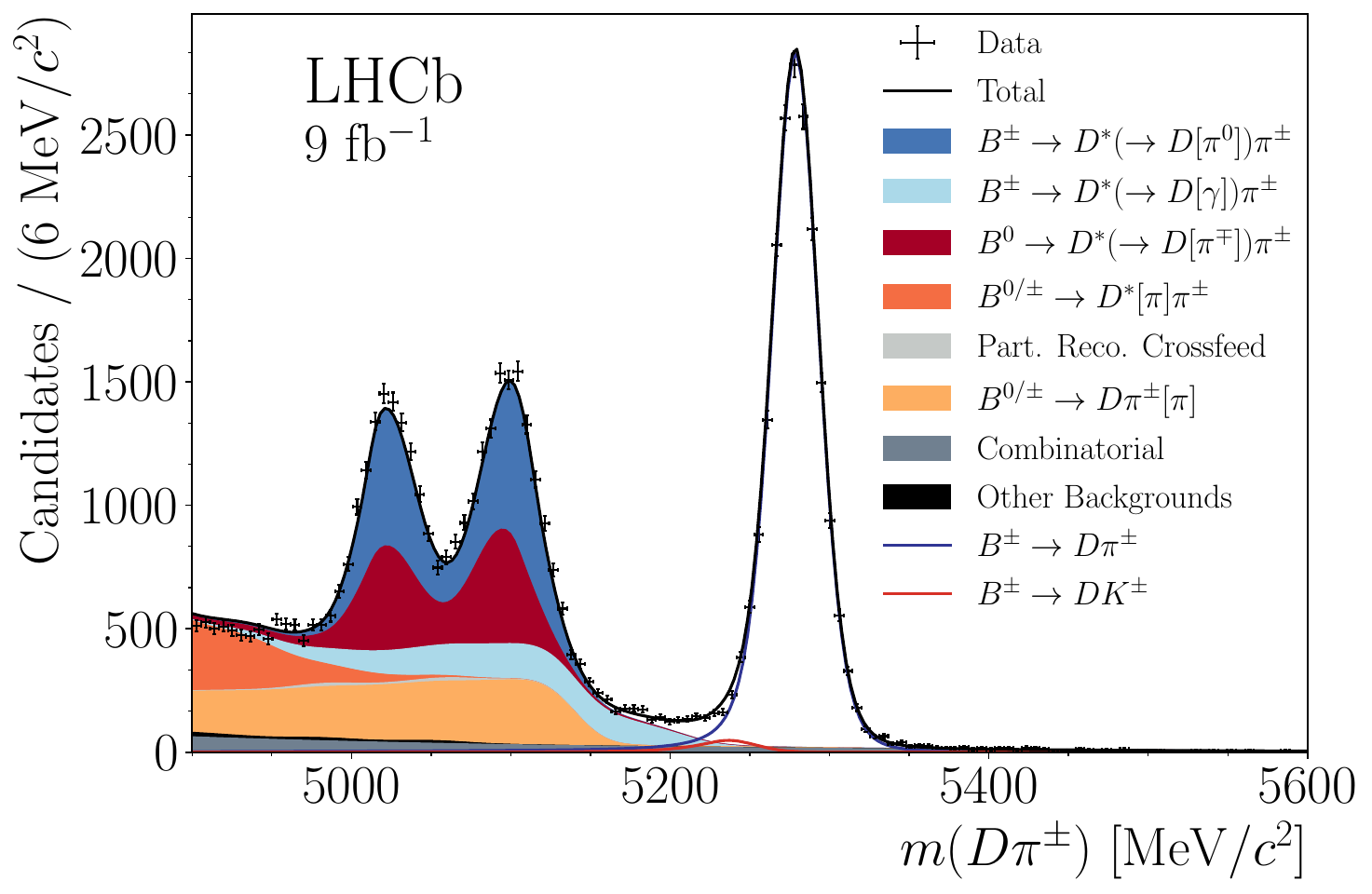}\\
    \includegraphics[width=0.48\textwidth]{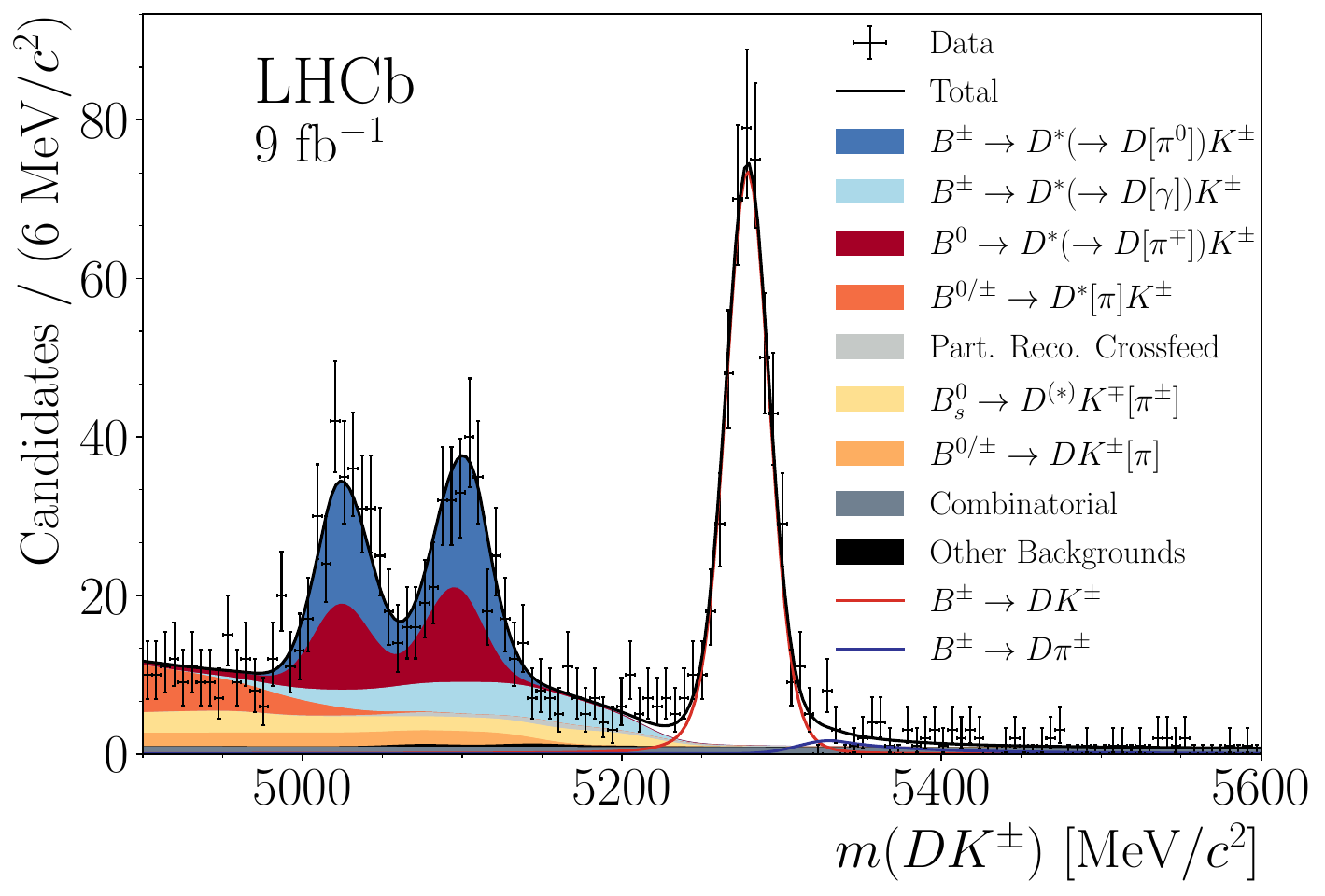} 
    \includegraphics[width=0.48\textwidth]{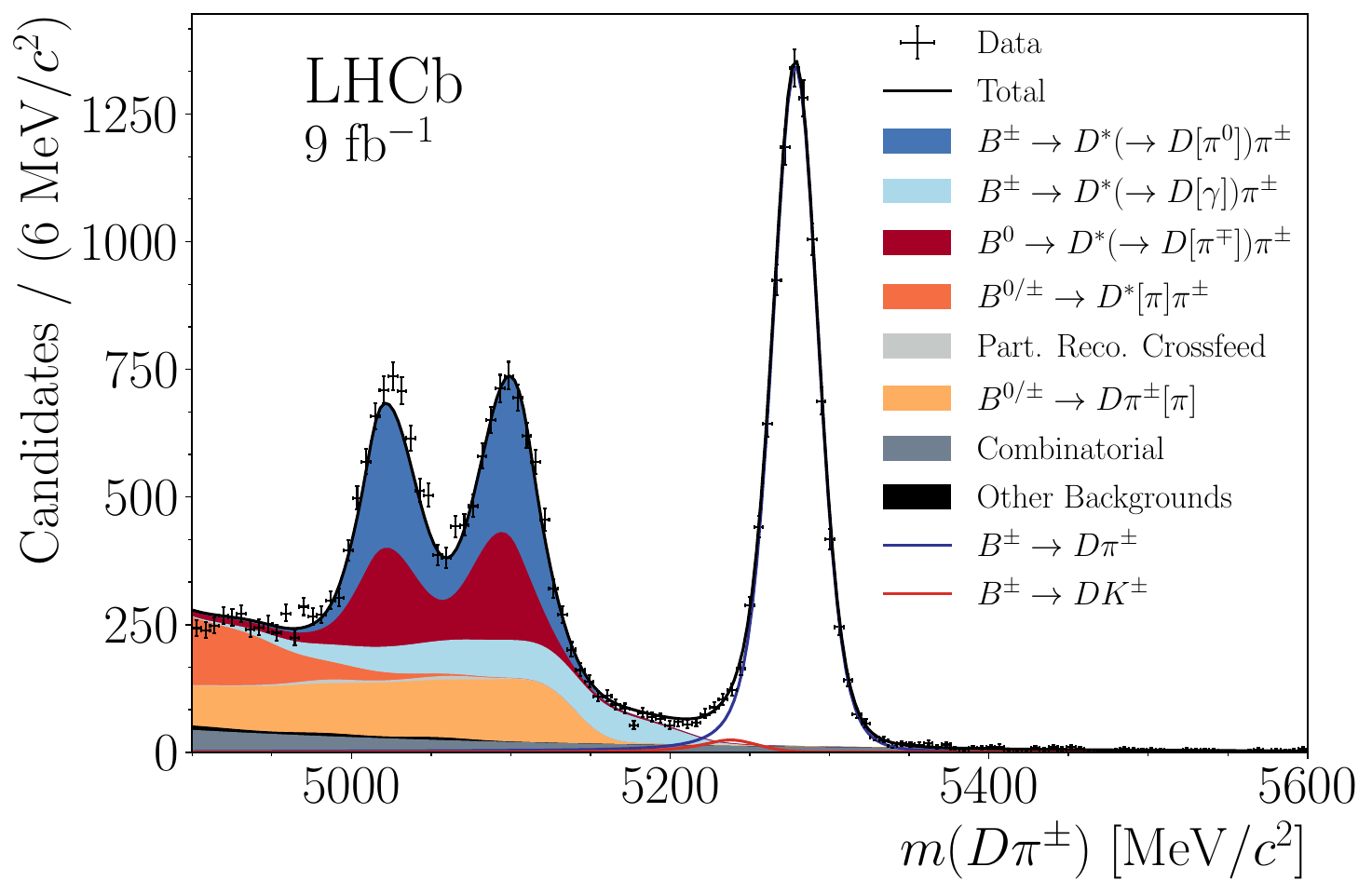}
    \caption{Mass distributions for (left) $\D\kaon$ samples and (right) $\D\pi$ samples, with $\D \to \KS \Kp \Km$ decays reconstructed with a (top) downstream \KS candidate and a (bottom) long \KS candidate. The projections of the fit results are overlaid. In the legend, particles in square brackets are not reconstructed.}
    \label{fig:globalfit_results_kskk}
\end{figure}

The results of the global fit are shown in Figs.~\ref{fig:globalfit_results_kspp} and~\ref{fig:globalfit_results_kskk} with signal and background yields over the fit region, 4900--5600 \mevcc, quoted in Table~\ref{tab:global_yields}. For most components, the uncertainty on their yield is smaller than their Poisson uncertainty. These are components where the ratio of yields has been fixed from external measurements or simulation. The yields of the fully reconstructed $\Bpm \to \D \hadron^{\pm}$ components can be compared to those in Ref.~\cite{LHCb-PAPER-2020-019} where these decays were treated as signal, and are found to be consistent considering selection differences. 

\sisetup{separate-uncertainty,table-align-uncertainty,table-format = 0(0)}

    \begin{table}[tb]
    \small
    \centering
        \caption{ Signal and background yields from the global fit over the full mass range, ${4900\textendash5600\mevcc}$. All yields and uncertainties are rounded to the nearest integer and uncertainties of 1 mean one or fewer. Backgrounds that are not written explicitly in this table are included in the `Other Backgrounds' component. }
\begin{tabular}{ l l | S S}

\hline
  &  & \multicolumn{2}{l}{Reconstructed as:}  \\
  $D$ decay & Component & $\Bpm \to \D\Kpm$& $\Bpm \to \D\pipm$ \\ 
\hline

$\D \to \KS \pip\pim$ & $\Bpm \to \Dstar[\D\piz]\Kpm$ & 6244 $~\pm~$ 12 & 2716 $~\pm~$ 5   \\ 
                        & $\Bpm \to \Dstar[\D\piz]\pipm$ & 340 $~\pm~$ 1 &  113170 $~\pm~$ 229 \\ 
                        & $\Bpm \to \Dstar[\D\g]\Kpm$ & 3144 $~\pm~$ 6 & 1247 $~\pm~$ 2 \\ 
                        & $\Bpm \to \Dstar[\D\g]\pipm$ & 166 $~\pm~$ 1 &  60285 $~\pm~$ 121  \\ 
                        & $\Bpm \to \D\Kpm$ & 10398 $~\pm~$ 21 & 4726 $~\pm~$ 9  \\ 
                        & $\Bpm \to \D\pipm$ & 590 $~\pm~$ 1 & 196804 $~\pm~$ 398  \\ 
                        & Other backgrounds & 10402 $~\pm~$ 105 & 206664 $~\pm~$ 592  \\ 
                        & Combinatorial background & 1343 $~\pm~$ 147 & 15177 $~\pm~$ 706\\ 
\hline
$\D \to \KS K^+ K^-$  & $\Bpm \to \Dstar[D\piz]\Kpm$ & 790 $~\pm~$ 3 & 344 $~\pm~$ 1 \\ 
                        & $\Bpm \to \Dstar[D\piz]\pipm$ & 43 $~\pm~$ 1 & 14327 $~\pm~$ 65 \\ 
                        & $\Bpm \to \Dstar[D\g]\Kpm$ & 397 $~\pm~$ 1 &  157 $~\pm~$ 1  \\ 
                        & $\Bpm \to \Dstar[D\g]\pipm$ & 21 $~\pm~$ 1 &  7636 $~\pm~$ 34\\ 
                        & $\Bpm \to \D\Kpm$ & 1527 $~\pm~$ 6 &  694 $~\pm~$ 2 \\ 
                        & $\Bpm \to \D\pi^\pm$ & 88 $~\pm~$ 1 & 29786 $~\pm~$ 135 \\ 
                        & Other backgrounds & 1573 $~\pm~$ 15 & 31278 $~\pm~$ 115 \\ 
                        & Combinatorial background & 263 $~\pm~$ 46 & 4413 $~\pm~$ 261 \\ 
\hline

\end{tabular}
    \label{tab:global_yields}
\end{table}

\section{Determination of the $\boldsymbol{\CP}$-violating observables}
\label{sec:CPFit}
 At this stage, a binned extended maximum-likelihood fit is performed to determine the \CP-violating observables. The categories are additionally split by \B meson charge and Dalitz bins. It is performed in the same mass range as the global fit from which all shape parameters are fixed. 

Different components have different distributions over the Dalitz plot depending on how they decay and whether the decay is \CP-violating. The signal is described using the set of equations shown in Eq.~\ref{eq:SigYieldEqns} where the \CP-violating observables, $x_{\pm}$ and $y_{\pm}$, and the normalisation factors are free in the fit. The same is true for the $\Bpm \to \D \hadron^{\pm}$ background where corresponding \CP-violating observables and normalisation factors are free in the fit. For other backgrounds originating from $B$ hadrons the integrated yield over the phase space is fixed from the global fit. Some of these background components are \CP-violating and are described using similar equations to Eq.~\ref{eq:SigYieldEqns} but with the addition of a coherence factor, $\kappa$, diluting the interference term, for example
\begin{equation}
    N^{+}_{i} \propto [F_{- i} + (x_{+}^{2} + y_{+}^{2})F_{+ i} + 2\kappa \sqrt{F_{- i}F_{+ i}}(c_{i}x_{+} - s_{i}y_{+})].
    \label{eq:BkgYieldEqns}
\end{equation}
Since these background contributions include multiple resonances, the coherence factor is not necessarily unity. These \CP-violating background components are parameterised with their own set of \CP-violating observables, which are fixed using hadronic parameters from Ref.~\cite{LHCb-CONF-2022-003} unless stated otherwise. For the $\Bz \to \D \Kpm \pimp$ and $\Bz \to \Dstar \Kpm \pimp$ background decays, $r_{B}$ values of 0.25 are used as this corresponds to that of the $\Bz \to \D \Kstarz$ decay. The same resonance is used to fix the strong phase for $\Bz \to \D \Kpm \pimp$ decays to 197\degrees. For the background from $\Bz \to \Dstar \Kpm \pimp$ decays a strong phase of 70\degrees is chosen from auxiliary studies of $\Bz \to \D \Kstarz$ used in Ref.~\cite{LHCB-CONF-2023-003}, while for those from $\Bpm \to D^{(*)} \Kpm \piz$ decays the $\Bpm \to \D \Kstarpm$ decay is used to fix the $r_{B}$ values to 0.106 and the strong phases at 35\degrees. For the $\Bpm \to D^{(*)} \pipm \piz$ backgrounds, $\Bpm \to \D \pipm$ decays are used to fix the $r_{B}$ values at 0.0048. The strong phases are arbitrarily chosen to be 100\degrees for the $\Bpm \to D \pipm \piz$ background and 200\degrees for the $\Bpm \to D^{*} \pipm \piz$ background decay since their values are unknown. The full range of values is used to determine the associated systematic uncertainty. The value of $\kappa$ is not well known for any of these decays. Hence, the central value is estimated using a single decay, by calculating the fraction of the $\Bz \to \D \Kstarz$ decay in the $\Bz \to \D \Kpm \pimp$ background contribution using branching fractions and efficiencies. The result is 0.5, and it is used as the nominal coherence factor for all the \CP-violating background components, except the $\Bpm \to \D \pipm \piz$ background decay. Here, it is well established that the background is dominated by the $\Bpm \to \D \rhopm$ resonant process~\cite{LHCb-PAPER-2015-059} and therefore a coherence factor of one is used. The systematic uncertainty associated with all these choices is determined by considering the full spectrum of reasonable variations in all hadronic parameters.

There are also \CP-conserving background components, which can exhibit one of three types of Dalitz-plot distributions. If the background decays via a $\Dz$$(\Dzb)$ and is reconstructed as a $B^{-}(B^{+})$ decay, it has the Dalitz-plot distribution of a \Dz meson. In this case the yield per Dalitz bin for the \Bpm decay, $N^{\pm}_{i}$, is proportional to $F_{\mp i}$. Background decays that fall into this category are $\Bz \to \Dstarm h^{+}$ and $\Bz \to [\D \pim]_{\Dstarm} \Kp \piz$ decays. Conversely, if the background decays via a $\Dzb(\Dz)$ and is assigned as a $B^{-}(B^{+})$ decay, it has the Dalitz-plot distribution of a \Dzb meson. Here the yield per Dalitz bin for the \Bpm decay, $N^{\pm}_{i}$, is proportional to $F_{\pm i}$. The $\Bs \to D^{(*)}\Km \pip$ and $\Bm \to \D \pim \pim \pip$ decays where pions with same-sign charges are not reconstructed fall into this category. Lastly, there are backgrounds which decay via either a \Dz or \Dzb with equal probability. The yield per Dalitz bin is proportional to the average of the $F_{+i}$ and $F_{-i}$ values. The $\Bz \to D^{(*)} \pip \pim$ decays fall into this category. The yields of the associated cross-feed components are determined with the same parameterisation used in the global fit. The distribution of the combinatorial background is not known and hence the yield varies freely in every Dalitz bin.

The \CP-violating backgrounds are parameterised similarly to Eq.~\ref{eq:SigYieldEqns} with fixed values of $x_{\pm}$ and $y_{\pm}$ that have large associated uncertainties. To avoid contamination of the signal measurements, two sets of $F_{i}$ values are introduced, one is used to parameterise the signal components, and another for all backgrounds, except the combinatorial background which has no dependence on $F_{i}$. Additionally, separate $F_{i}$ values are used for long and downstream \KS candidates. The $F_{i}$ parameters are implemented in the fit in terms of recursive fractions, identically to Ref.~\cite{LHCb-PAPER-2020-019}, and vary freely in the \CP fit. For the signal $F_{i}$ parameters, the values are driven by the $\Bpm \to \Dstar \pipm$ component, and for the background $F_{i}$ parameters the dominant contribution comes from the $\Bpm \to \D\pipm$ and $\Bpm \to \D\rhopm$ components.

In summary, in the \CP fit the parameters that are allowed to vary are the two sets of six cartesian \CP-violating observables for the signal and $\Bpm \to \D \hadron^{\pm}$ decays, two sets of $F_{i}$ parameters, the yields of the combinatorial background in each Dalitz bin, and the normalisation factors given in Eq.~\ref{eq:SigYieldEqns}.

Instabilities and biases in the fit are studied using pseudoexperiments that are generated and fitted using the baseline fit model. All \CP-violating observables show normalised residuals consistent with a normal Gaussian distribution except the $\Re(\xi^{\Dstar\pi})$ parameter that exhibits a bias of 15\% and a 6\% over-coverage in its statistical uncertainty. Both the bias and the overestimated uncertainty are found to be due to statistical fluctuations leading to small data yields in certain bins and are reduced if a larger data sample is generated. The measured $\Re(\xi^{\Dstar\pi})$ parameter is corrected for both of these features. The \CP-violating observables for the $\Bpm \to \D \hadron^{\pm}$ decays are cross-checked against results from Ref.~\cite{LHCb-PAPER-2020-019} and are found to be consistent. 

\begin{figure}[tb]
\includegraphics[width=\textwidth]{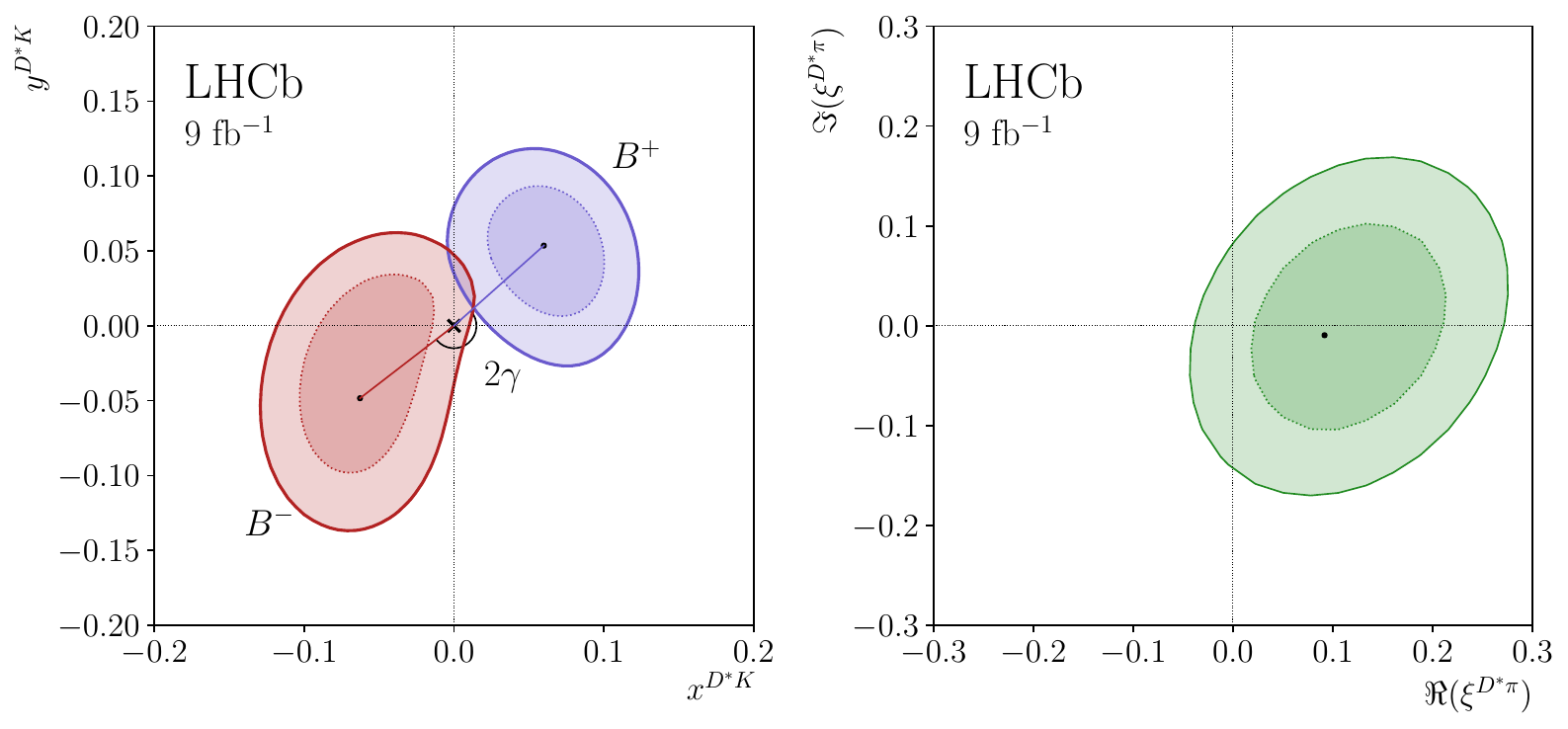}
    \caption{Contours at 68.3\% and 95.5\% confidence levels indicated by the darker and lighter regions, respectively, for (left, red) $(x_{-}^{\Dstar \kaon}, y_{-}^{\Dstar \kaon})$ which is labelled $\Bm$ and (left, purple) $(x_{+}^{\Dstar \kaon}, y_{+}^{\Dstar \kaon})$ which is labelled $\Bp$, and (right) $(\Re(\xi^{\Dstar\pi}), \Im(\xi^{\Dstar \pi}))$, including only statistical uncertainty. The black points indicate the central values.}
    \label{fig:CPObs}
\end{figure}

The results of the \CP-violating observables for the signal are shown in Fig.~\ref{fig:CPObs}. The 68.3\% and 95.5\% confidence regions are determined using a likelihood scan and include only statistical uncertainties. In the left-hand plot of Fig.~\ref{fig:CPObs}, the angle between the two lines converging at the origin is 2\g, and the length of each line is $r_{B}^{\Dstar\kaon}$. In the right-hand plot of Fig.~\ref{fig:CPObs}, the \CP-violating observables for the $\Bpm \to \Dstar \pipm$ decay is shown, where the distance from the origin to the central value point is the ratio $r_{B}^{\Dstar \pi}/r_{B}^{\Dstar \kaon}$, which is around 0.1 as expected~\cite{LHCb-CONF-2022-003}.

\section{Systematic uncertainties}
\label{sec:Systematics}
The various sources of systematic uncertainty are summarised in Table~\ref{tab:syst}. A systematic uncertainty is assigned to account for the fact that the \D decay phase-space efficiency is assumed to be uniform. The strong-phase inputs are based on measurements at \cleo and \besiii and have been corrected for efficiency effects at these charm factories. As a result, there is a small difference between the strong-phase inputs used in this paper and those that would be determined if the strong-phase inputs were modulated by $\eta(s)$. The systematic uncertainty is evaluated using amplitude models in Ref.~\cite{PhysRevD.98.112012} to calculate strong phases with and without efficiency corrections which are determined using simulation. Pseudoexperiments are then generated using strong phases with efficiency corrections and fitted using strong phases without efficiency corrections. The mean bias in each \CP-violating observable is taken as the associated systematic uncertainty.

There is a considerable systematic uncertainty due to the mass shape parameterisation. This refers to any fixed mass shape parameter in the fits to simulation or in the global fit. A bootstrapping method~\cite{efron:1979} is used to ensure statistical uncertainties from the global fit and correlations between shape parameters are accounted for in the error propagation. First, the simulation samples are re-sampled with replacement and the fits to simulation are repeated many times. For each fit result, data are re-sampled with replacement and the global and \CP fits are repeated. The standard deviation of each \CP-violating observable is an estimate of the associated systematic uncertainty.

There are a number of fixed quantities in the global and \CP fits. The systematic uncertainties for these are estimated by varying each fixed quantity within its uncertainty and repeating the fits. The standard deviation of the fitted \CP-violating observables is the associated systematic uncertainty. The fixed quantities that are varied include the asymmetry parameters for the $\Bpm \to [\D\g]_{\Dstar}h^{\pm}$ signal, $\xi_{\D\g}$, the branching fractions, the efficiencies, and the yield ratios from Ref.~\cite{LHCb-PAPER-2020-036}. The latter also includes an uncertainty associated with the assumption in the parameterisation of Eq.~\ref{eq:Bs_DKpi_yield_param} that the efficiencies in this analysis and Ref.~\cite{LHCb-PAPER-2020-036} are the same. Differences of up to 3\% are estimated with the simulation, and are included in the variations. 

A bias correction is applied to the \CP-violating observables from the \CP fit, which is determined using pseudoexperiments. The size of the biases can vary depending on the input parameters. To determine the systematic uncertainty associated with this the range of biases on the \CP-violating observables are assessed by testing alternative inputs.

There is also a systematic uncertainty due to the resolution of the momentum measurements leading to event migration between neighbouring Dalitz bins. The effect is accounted for at first order in the $F_{i}$ parameters, but second order effects remain due to differing intensity distributions between the Dalitz plots for $\Bpm \to \Dstar \Kpm$ and $\Bpm \to \Dstar \pipm$ decays. The size of the systematic uncertainty is determined in the same way as in Ref.~\cite{LHCb-PAPER-2020-019} and found to be negligible.

There are a number of \CP-violating background contributions which are fixed in the \CP fit. Their inputs carry uncertainties, therefore a systematic uncertainty is assigned. It is evaluated by generating pseudoexperiments with alternative inputs and fitting to the baseline model. In two sets of studies, the hadronic parameters are varied to the upper and lower limits of their 68\% confidence levels given in Ref.~\cite{LHCb-CONF-2022-003}. Then, the larger mean bias in each \CP-violating observable from these two studies is taken as its systematic uncertainty. The coherence factors are varied conservatively as there is no prior knowledge for most background components. For the $\Bz \to \D \Kpm \pimp$ background decay the coherence factor is shifted to 0.958, the value measured for the $\Bz \to \D \Kstarz(892)$ decay in an amplitude measurement of the $\Bz \to \D \Kp \pim$ decay~\cite{LHCb-PAPER-2015-059}. For all other backgrounds a value of one is used, except for the $\Bpm \to \D \pipm \piz$ decay for which there is no variation.

The leading sources of experimental systematic uncertainties (mass shape parametrisation, fixed branching ratios, fixed yield ratios, and inputs for CPV backgrounds) are driven by the parameterisation of two background components, ${\Bpm \to \D\Kpm \piz}$ and ${\Bpm \to \D\pipm \piz}$ decays. Further direct study of these decays will provide the information necessary to reduce systematic uncertainties in a future version of the measurement described here.

The systematic uncertainty due to the external strong-phase inputs is evaluated by varying the $c_{i}$ and $s_{i}$ values within their measured uncertainties taking into account correlations between them and repeating the fits many times. The standard deviation of the resulting \CP-violating observables is assigned as the systematic uncertainty.

As determined in Ref.~\cite{LHCb-PAPER-2020-019} the systematic uncertainties associated with ignoring \CP violation and regeneration in \KS interactions with matter and charm mixing are expected to be minimal and are not evaluated. The total systematic uncertainty is found to be at least a factor of two smaller than the statistical uncertainty.

\begin{table}[tb]
    \centering
    \small
        \caption{Summary of the systematic uncertainties. Values are expressed in units of $10 ^{-2}$.}
        \begin{adjustbox}{width=0.9\textwidth}
{\renewcommand{\arraystretch}{1.2}
\begin{tabular}{ l | c c c c c c }

\hline

Source & $x_{-}^{D^*K}$ & $y_{-}^{D^*K}$ & $x_{+}^{D^*K}$ & $y_{+}^{D^*K}$ & $\Re(\xi^{D^*\pi})$ & $\Im(\xi^{D^*\pi})$ \\
\hline
Efficiency correction of $(c_{i}, s_{i})$ &  0.23 & 0.29 & 0.21 & 0.20 & 0.47 & 0.31 \\

Mass shape parameterisation & 0.35 & 0.58 & 0.38 & 0.33 & 1.17 & 0.90 \\

Fixed $\xi_{D\gamma}$ parameter & 0.14 & 0.19 & 0.15 & 0.08 & 0.22 & 0.32 \\

Fixed branching ratios & 0.58 & 0.44 & 0.33 & 0.50 & 1.09 & 0.54 \\

Fixed efficiencies & 0.23 & 0.48 & 0.18 & 0.27 & 0.70 & 0.38\\

Fixed yield ratios  & 0.66 & 0.85 & 0.46 & 0.43 & 1.45 & 0.77 \\

Bias Correction & 0.29 & 0.35 & 0.12 & 0.16 & 0.62 & 0.51 \\

Dalitz-bin migration &  0.00 & 0.02 & 0.04 & 0.10 & 0.03 & 0.11 \\

Inputs for CPV backgrounds & 0.35 & 0.33 & 0.38 & 0.21 & 2.22 & 1.93 \\

\hline
Total of above uncertainties  & 1.11 & 1.36 & 0.85 & 0.87 & 3.28 & 2.46 \\
\hline
Strong-phase inputs &  0.57 & 1.54 & 0.18 & 0.41 & 2.33 & 2.13\\
Total systematic uncertainty& 1.25 & 2.05 & 0.87 & 0.95 & 4.02 & 3.26\\
\hline
Statistical uncertainty&  2.93 & 5.69 & 2.58 & 2.87 & 9.37 & 9.67 \\
\hline
    \end{tabular}
}
    \end{adjustbox}
    \label{tab:syst}
\end{table}

\section{Interpretation}
\label{sec:Interpretation}
\begin{figure}[tb]
    \centering
    \includegraphics[width=0.6\textwidth]{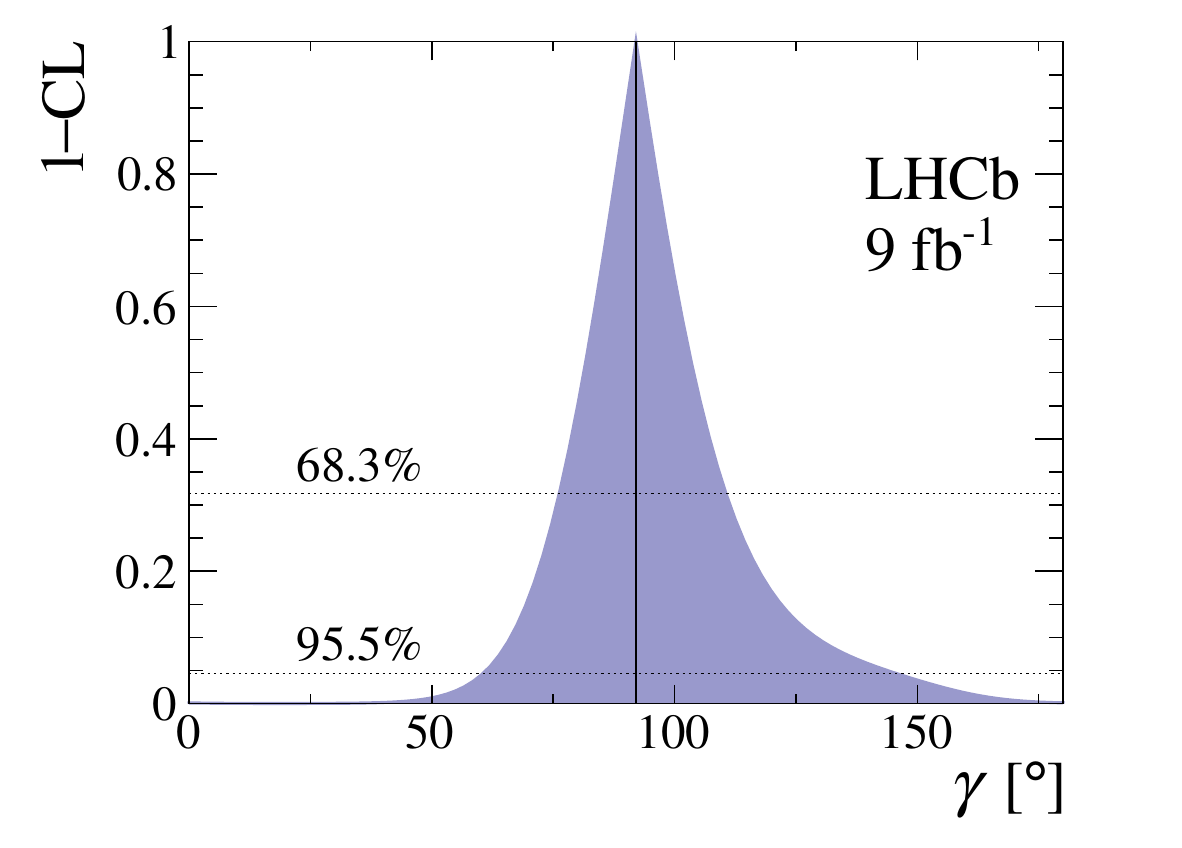}
    \caption{Confidence intervals at 68.3\% and 95.5\% for the CKM angle \g.}
    \label{fig:interp_g}
\end{figure}

\begin{figure}
    \centering
    \includegraphics[width=0.45\textwidth]{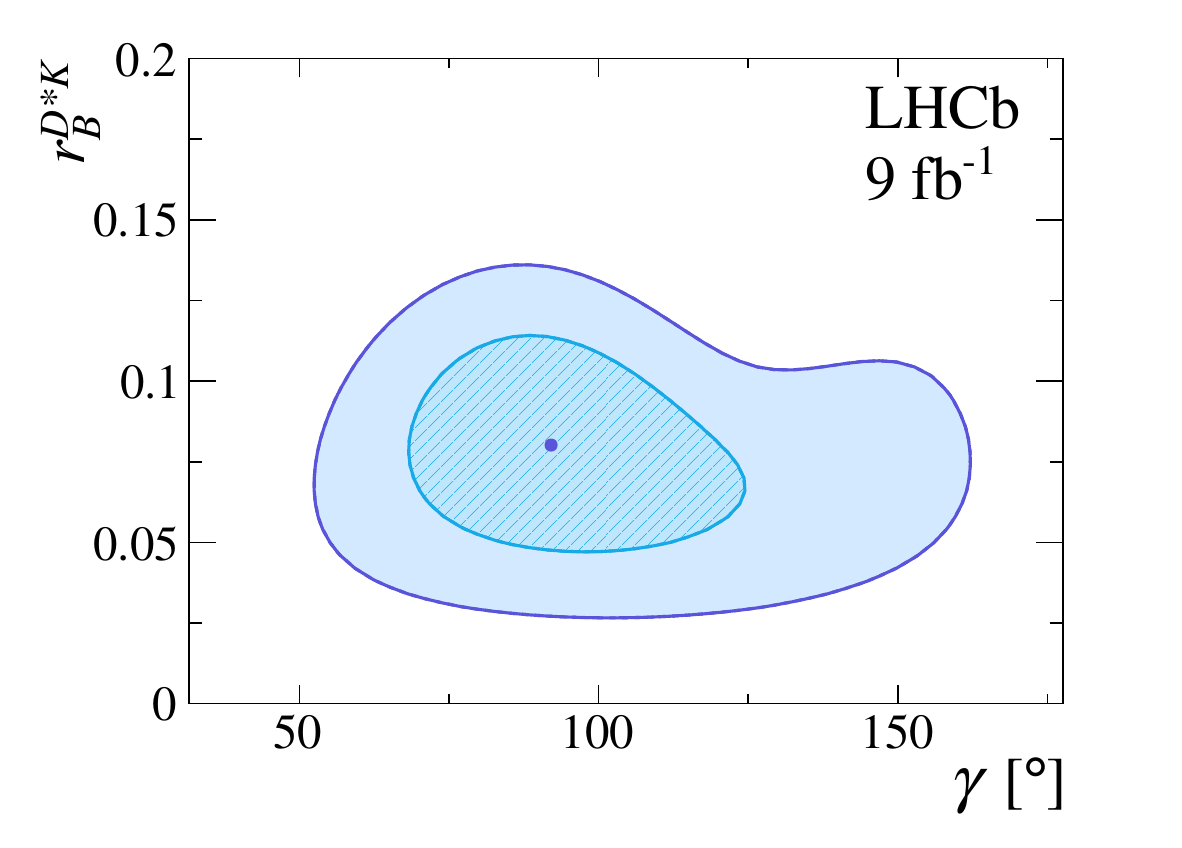}
    \includegraphics[width=0.45\textwidth]{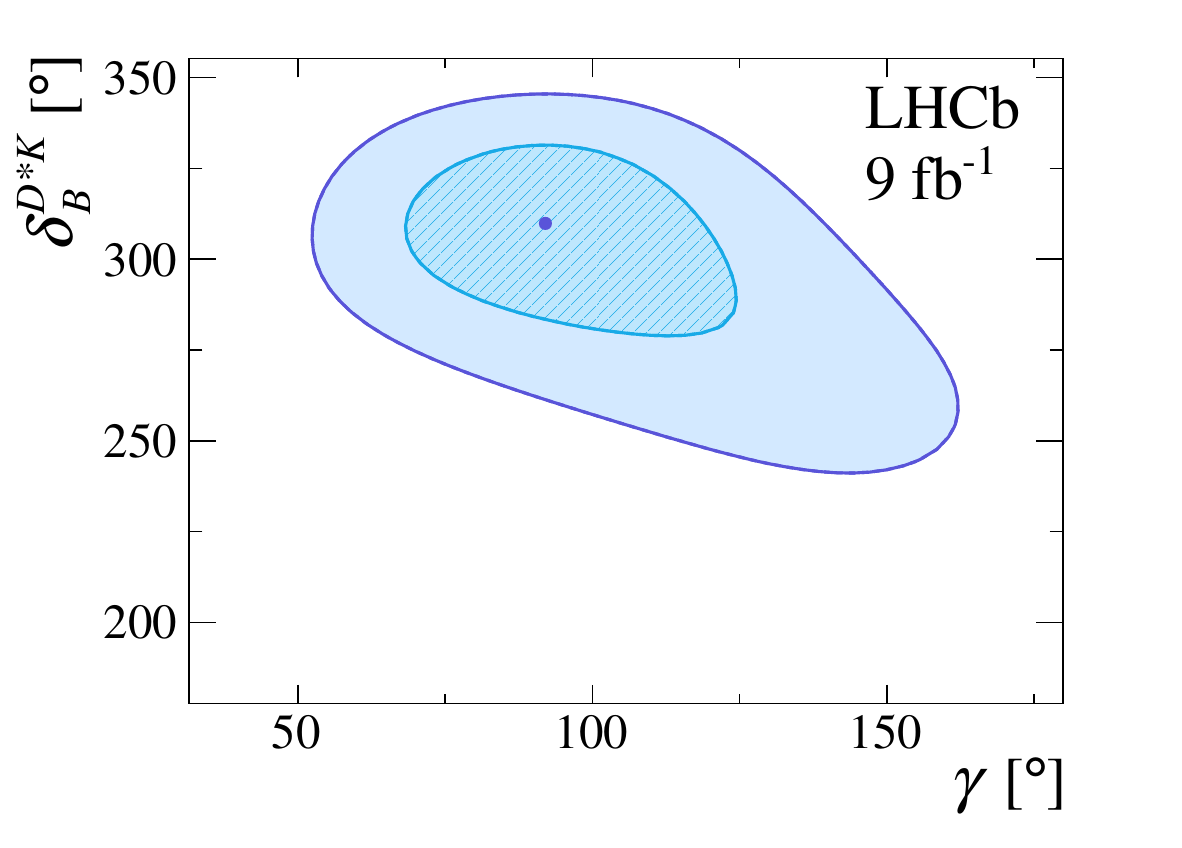}\\
    \includegraphics[width=0.45\textwidth]{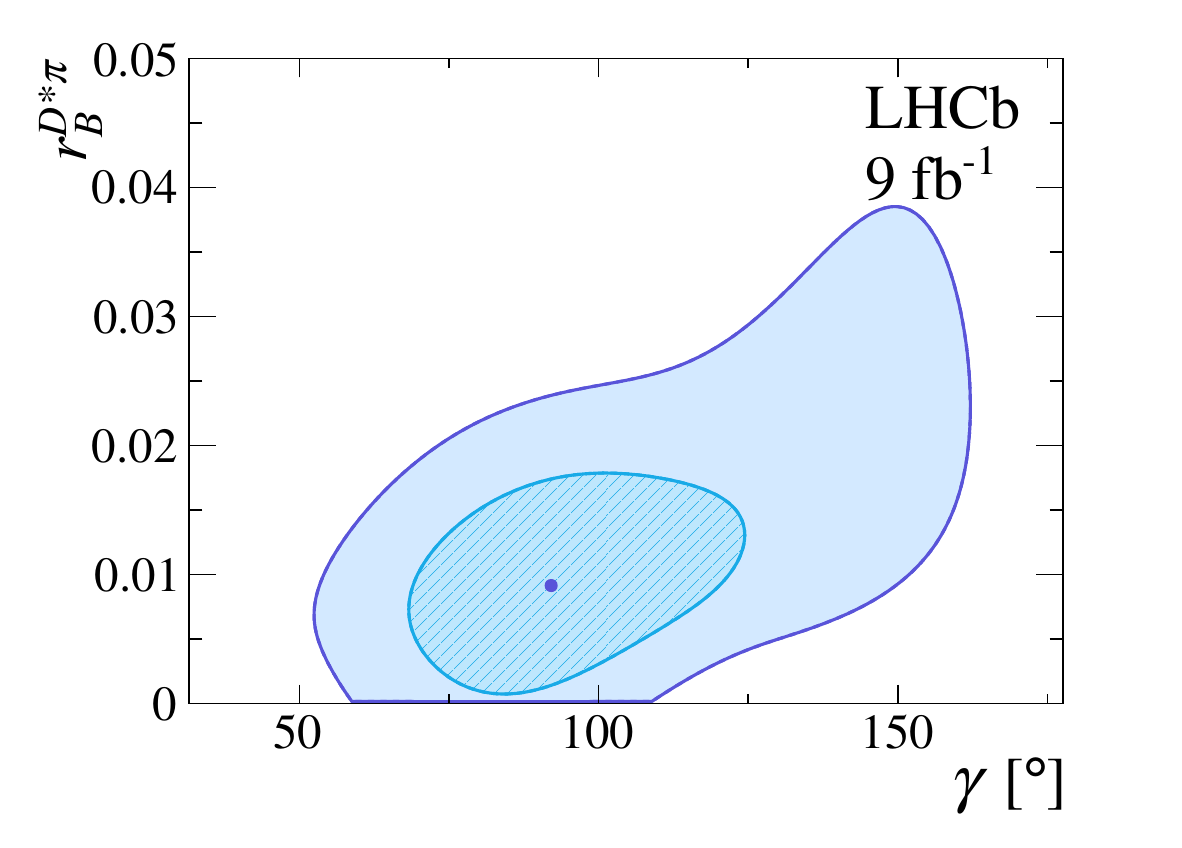}
    \includegraphics[width=0.45\textwidth]{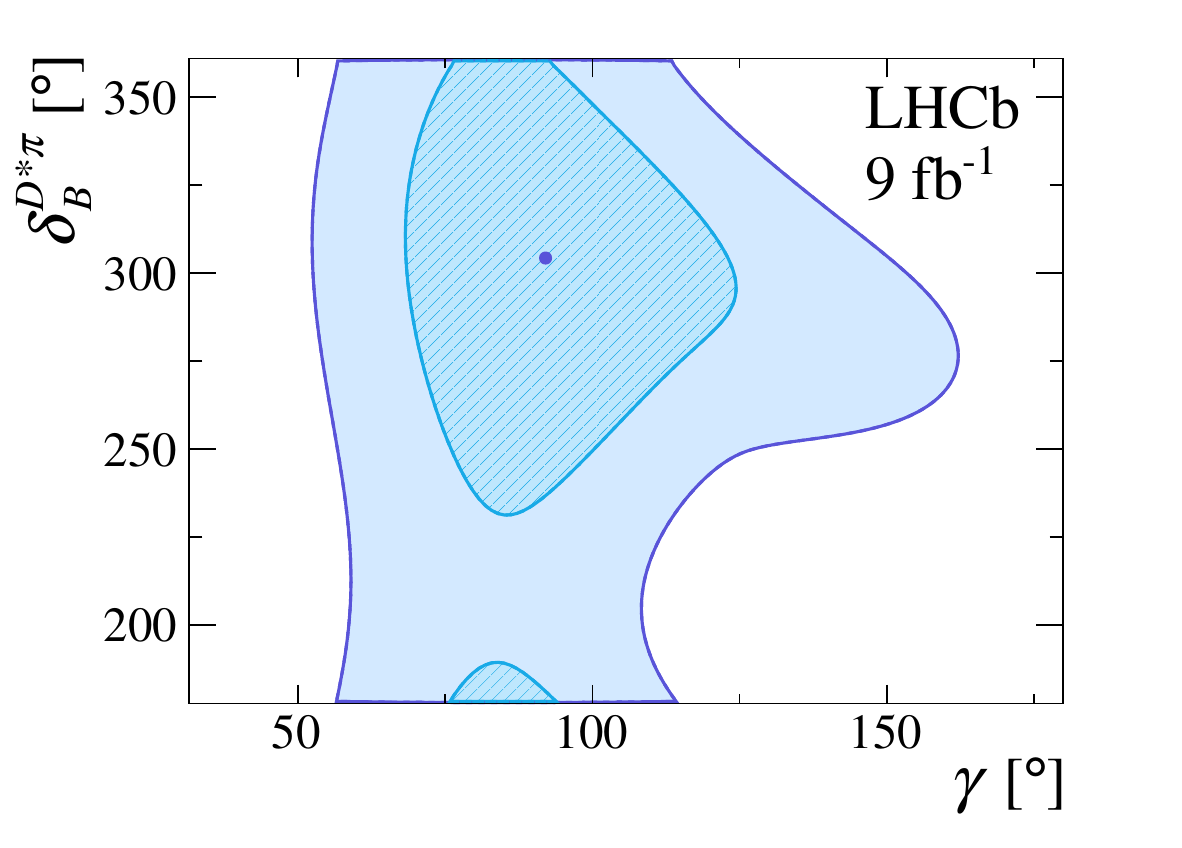}\\
    \caption{Confidence regions at 68.3\% and 95.5\% indicated by the lighter and darker shades respectively, for (top, left) \g vs. $r_{B}^{\Dstar\kaon}$, (top, right) \g vs. $\delta_{B}^{\Dstar\kaon}$, (bottom left) \g vs. $r_{B}^{\Dstar\pi}$, and (bottom, right) \g vs. $\delta_{B}^{\Dstar\pi}$.}
    \label{fig:2Dcontours}
\end{figure}

The \CP-violating observables are measured to be
\begin{align*}
x_{-}^{\Dstar \kaon} &=  ( -6.3  \pm 2.9 \pm 1.1 \pm 0.6) \times 10^{-2},\\
y_{-}^{\Dstar \kaon} &=  ( -4.8 \pm  5.7 \pm 1.4 \pm 1.5) \times 10^{-2}, \\
x_{+}^{\Dstar \kaon} &= ( \phantom{-} 6.0  \pm 2.6 \pm 0.9 \pm 0.2) \times 10^{-2}, \\
y_{+}^{\Dstar \kaon} &= ( \phantom{-}5.4  \pm 2.9  \pm 0.9 \pm 0.4) \times 10^{-2}, \\
\Re(\xi^{\Dstar \pi}) &= ( \phantom{l}11.5  \pm 9.4 \pm 3.3 \pm 2.3) \times 10^{-2}, \\
\Im(\xi^{\Dstar \pi}) &= ( -0.9  \pm 9.7 \pm 2.5 \pm 2.1) \times 10^{-2},
\end{align*}
where the first uncertainty is statistical, the second is systematic, and the third is due to external strong-phase inputs from \besiii~\cite{Ablikim_2020pi, Ablikim_2020K} and \cleo~\cite{Libby_2010}. The correlation matrices for each of these uncertainties are given in Tables~\ref{tab:stat_corr}--\ref{tab:cisi_corr} in Appendix~\ref{app:corr}. The \CP-violating observables are interpreted in terms of the physical parameters, $r_{B}^{\Dstar\kaon}$, $\delta_{B}^{\Dstar\kaon}$, $r_{B}^{\Dstar\pi}$, $\delta_{B}^{\Dstar\pi}$, and $\gamma$ using a maximum-likelihood fit with a frequentist approach~\cite{LHCb-PAPER-2016-032}. The 68.3\% and 95.5\% confidence regions are shown in Figs.~\ref{fig:interp_g} and~\ref{fig:2Dcontours}. The parameterisation in Eq.~\ref{eq:CPobs_def} allows a two-fold symmetry with solutions where $\g \to \g + 180\degrees$ and $\delta_{B}\to \delta_{B} + 180\degrees$. Choosing $0 < \g < 180\degrees$, the numerical solutions are
\begin{align*}
    \g &= (92^{+21}_{-17})\degrees,\\
    r_{B}^{\Dstar K} &= 0.080^{+0.022}_{-0.023},\\
    \delta_{B}^{\Dstar K} &= (310^{+15}_{-20})\degrees,\\
    r_{B}^{\Dstar \pi} &= 0.009^{+0.005}_{-0.007},\\
    \delta_{B}^{\Dstar \pi} &= (304^{+37}_{-38})\degrees.
\end{align*}
The solution for \g is consistent with the latest \g combination using \lhcb data, ${\g=(63.8^{+3.5}_{-3.7})\degrees}$~\cite{LHCb-CONF-2022-003}.

In order to compare results from this analysis and Ref.~\cite{LHCb-PAPER-2023-012}, it is necessary to ascertain the level of statistical correlation between them. All selected events in Ref.~\cite{LHCb-PAPER-2023-012} also appear in this analysis. The analysis in Ref.~\cite{LHCb-PAPER-2023-012} has 45\% (18\%) of the $\Bpm \to [\D\g]_{\Dstar}\Kpm$ $(\Bpm \to [\D\piz]_{\Dstar}\Kpm)$ signal yield compared to this analysis. However, a number of differences between the two analyses mean common candidates do not carry the same weight. First, different \Dstar decays drive the sensitivity of each analysis. Furthermore, in Ref.~\cite{LHCb-PAPER-2023-012}, the two signal channels are well separated, whereas here, both decays occupy the same $m(\D\hadron)$ region and dilute each others' sensitivities. Finally, the background environments differ between the two analyses. These differences are incorporated into studies with pseudoexperiments which are used to estimate the statistical correlation. It is determined to be a maximum of 3\%, and can therefore be treated as negligible. It is noted that the experimental systematic uncertainties are uncorrelated and those of the external inputs are determined.

The \CP-violating observables determined in this analysis and Ref.~\cite{LHCb-PAPER-2023-012} are compared and found to be consistent. For the $x_\pm^{\Dstar\kaon}$ and $y_\pm^{\Dstar\kaon}$ observables this analysis has overall uncertainties that are 30\% smaller, except for $y_{-}^{\Dstar\kaon}$ where it is 30$\%$ larger. The uncertainties on the $\xi^{\Dstar \pi}$ parameter are almost twice as large as those in Ref.~\cite{LHCb-PAPER-2023-012} due to the dominant presence of the $\Bpm \to \D \rhopm$ background which is suppressed when the \Dstar decay is reconstructed. In Ref.~\cite{LHCb-PAPER-2023-012}, the uncertainty on \g is smaller at $14\degrees$ where the sensitivity to \g is enhanced by the corresponding larger value of $r_{B}^{\Dstar K}$. The central values can also be compared to the $\Bpm \to \Dstar \Kpm$, $\D \to hh$ results in Ref.~\cite{LHCb-PAPER-2020-036}, and are also found to be consistent. The measurement presented in this paper adds further information on \CP violation in $\Bpm \to \Dstar \hadron^{\pm}$ decays and, in combination with other measurements of the same $B$ decay mode~\cite{LHCb-PAPER-2023-012,LHCb-PAPER-2020-036}, it will set strong constraints on the CKM angle \g and associated hadronic parameters by removing the ambiguities resulting from multiple solutions in the results of Ref.~\cite{LHCb-PAPER-2020-036}.

\section{Conclusion}
\label{sec:Conclusion}
In this work $\Bpm \to \Dstar \Kpm$ and $\Bpm \to \Dstar \pipm$ decays with partially reconstructed ${\Dstar \to \D\g}$ and ${\Dstar \to \D\piz}$ decays, followed by ${\D \to \KS \pip \pim}$ and ${\D \to \KS \Kp \Km}$ decays are studied at the \lhcb experiment. The full \lhcb Run 1 and Run 2 dataset is used to determine the CKM angle \g. The measurement is performed in Dalitz bins of the \D decay phase space in a model-independent manner, using \cleo and \besiii inputs for \D decay strong-phase information. A measurement of the CKM angle $\g=(92^{+21}_{-17})\degrees$ is achieved where the dominant contribution to the uncertainty is statistical.

% Do not include this in any draft (just for information in the template)
\section*{Acknowledgements}
%
% These Acknowledgements valid from 3-May-2019
%
\noindent We express our gratitude to our colleagues in the CERN
accelerator departments for the excellent performance of the LHC. We
thank the technical and administrative staff at the LHCb
institutes.
We acknowledge support from CERN and from the national agencies:
CAPES, CNPq, FAPERJ and FINEP (Brazil); 
MOST and NSFC (China); 
CNRS/IN2P3 (France); 
BMBF, DFG and MPG (Germany); 
INFN (Italy); 
NWO (Netherlands); 
MNiSW and NCN (Poland); 
MCID/IFA (Romania); 
%MSHE (Russia); 
MICINN (Spain); 
SNSF and SER (Switzerland); 
NASU (Ukraine); 
STFC (United Kingdom); 
DOE NP and NSF (USA).
We acknowledge the computing resources that are provided by CERN, IN2P3
(France), KIT and DESY (Germany), INFN (Italy), SURF (Netherlands),
PIC (Spain), GridPP (United Kingdom), 
%RRCKI and Yandex LLC (Russia), 
CSCS (Switzerland), IFIN-HH (Romania), CBPF (Brazil),
and Polish WLCG (Poland).
We are indebted to the communities behind the multiple open-source
software packages on which we depend.
Individual groups or members have received support from
ARC and ARDC (Australia);
Key Research Program of Frontier Sciences of CAS, CAS PIFI, CAS CCEPP, 
Fundamental Research Funds for the Central Universities, 
and Sci. \& Tech. Program of Guangzhou (China);
Minciencias (Colombia);
EPLANET, Marie Sk\l{}odowska-Curie Actions, ERC and NextGenerationEU (European Union);
A*MIDEX, ANR, IPhU and Labex P2IO, and R\'{e}gion Auvergne-Rh\^{o}ne-Alpes (France);
%RFBR, RSF and Yandex LLC (Russia);
AvH Foundation (Germany);
ICSC (Italy); 
GVA, XuntaGal, GENCAT, Inditex, InTalent and Prog.~Atracci\'on Talento, CM (Spain);
SRC (Sweden);
the Leverhulme Trust, the Royal Society
 and UKRI (United Kingdom).

\appendix

\section{Correlation Matrices}
Table~\ref{tab:stat_corr} provides the correlation matrix for the statistical uncertainties of the \CP-violating observables. Tables~\ref{tab:lhcb_syst_corr} and \ref{tab:cisi_corr} provide the correlations between the experimental systematic uncertainties and those arising from external strong-phase information, respectively.
\label{app:corr}

\begin{table}[tb]
    \centering
        \small
        \caption{Statistical correlation matrix for the \CP-violating observables.}
{\renewcommand{\arraystretch}{1.2}
\begin{tabular}{ c | c c c c c c}

& $x_{-}^{\Dstar\kaon}$ & $y_{-}^{\Dstar\kaon}$ & $x_{+}^{\Dstar\kaon}$ & $y_{+}^{\Dstar\kaon}$ & $\Re(\xi^{\Dstar\pi})$ & $\Im(\xi^{\Dstar\pi})$ \\
 \hline
 $x_{-}^{\Dstar\kaon}$ &   1.000 &       0.420 &   0.158  &  $\phantom{-}0.105$   &  0.445 &   0.422\\
 $y_{-}^{\Dstar\kaon}$ & &   1.000   &     0.115 &   $\phantom{-}0.232$  &  0.765 &   0.631\\
 $x_{+}^{\Dstar\kaon}$  & & &   1.000     &  $-0.095$  & 0.012 &   0.409\\
 $y_{+}^{\Dstar\kaon}$ & & & &    $\phantom{-}1.000$  &     0.263&    0.112\\
 $\Re(\xi^{\Dstar\pi})$ & & & & &   1.000 &       0.597\\
 $\Im(\xi^{\Dstar\pi})$ & & & & & &   1.000 \\
    \end{tabular}
}
    \label{tab:stat_corr}
\end{table}

\begin{table}[tb]
    \centering
        \small
        \caption{Correlation matrix associated with LHCb-related systematic uncertainties.}
{\renewcommand{\arraystretch}{1.2}
\begin{tabular}{ l | c c c c c c }

 & $x_{-}^{\Dstar\kaon}$ & $y_{-}^{\Dstar\kaon}$ & $x_{+}^{\Dstar\kaon}$ & $y_{+}^{\Dstar\kaon}$ & $\Re(\xi^{\Dstar\pi})$ & $\Im(\xi^{\Dstar\pi})$ \\

\hline
$x_{-}^{\Dstar\kaon}$ & 1.000 & 0.630 & $-0.241$ & $-0.016$ & 0.602 & $\phantom{-}0.083$ \\
$y_{-}^{\Dstar\kaon}$ & & 1.000 & $\phantom{-}0.008$ & $\phantom{-}0.154$ & 0.735 & $\phantom{-}0.230$ \\
$x_{+}^{\Dstar\kaon}$ & & & $\phantom{-}1.000$ & $\phantom{-}0.515$ & 0.232 & $\phantom{-}0.618$\\
$y_{+}^{\Dstar\kaon}$ & & & & $\phantom{-}1.000$ & 0.237 & $\phantom{-}0.112$ \\
$\Re(\xi^{\Dstar\pi})$ & & & & & 1.000 & $-0.201$\\
$\Im(\xi^{\Dstar\pi})$ & & & & & & $\phantom{-}1.000$ \\
    \end{tabular}
}
    \label{tab:lhcb_syst_corr}
\end{table}

\begin{table}[tb]
    \centering
    \small
        \caption{Correlation matrix associated with the strong-phase inputs.}
{\renewcommand{\arraystretch}{1.2}
\begin{tabular}{ l | c c c c c c }

 & $x_{-}^{\Dstar\kaon}$ & $y_{-}^{\Dstar\kaon}$ & $x_{+}^{\Dstar\kaon}$ & $y_{+}^{\Dstar\kaon}$ & $\Re(\xi^{\Dstar\pi})$ & $\Im(\xi^{\Dstar\pi})$ \\

\hline
$x_{-}^{\Dstar\kaon}$ & 1.000 & 0.886 & $-0.117$ & $-0.063$ & $\phantom{-}0.916$ & $\phantom{-}0.904$ \\
$y_{-}^{\Dstar\kaon}$ & & 1.000 & $-0.124$ & $-0.115$ & $\phantom{-}0.935$ & $\phantom{-}0.889$ \\
$x_{+}^{\Dstar\kaon}$ & & & $\phantom{-}1.000$ & $-0.280$ & $-0.130$ & $\phantom{-}0.063$\\
$y_{+}^{\Dstar\kaon}$ & & & & $\phantom{-}1.000$ & $\phantom{-}0.085$ & $-0.077$ \\
$\Re(\xi^{\Dstar\pi})$ & & & & & $\phantom{-}1.000$ & $\phantom{-}0.941$\\
$\Im(\xi^{\Dstar\pi})$ & & & & & & $\phantom{-}1.000$ \\
    \end{tabular}
}
    \label{tab:cisi_corr}
\end{table}

% This should be taken out in the final paper
\clearpage

% \section{Supplementary material for LHCb-PAPER-20XX-YYY}
% \label{sec:Supplementary-App}

\clearpage

\addcontentsline{toc}{section}{References}
%\setboolean{inbibliography}{true}
\bibliographystyle{LHCb}
\bibliography{main,standard,LHCb-PAPER,LHCb-CONF,LHCb-DP,LHCb-TDR}

\newpage
% LHCb collaboration author list
% Data extracted on November 13th, 2023 at 12:45pm for paper reference LHCb-PAPER-2023-029
\centerline
{\large\bf LHCb collaboration}
\begin
{flushleft}
\small
R.~Aaij$^{35}$\lhcborcid{0000-0003-0533-1952},
A.S.W.~Abdelmotteleb$^{54}$\lhcborcid{0000-0001-7905-0542},
C.~Abellan~Beteta$^{48}$,
F.~Abudin{\'e}n$^{54}$\lhcborcid{0000-0002-6737-3528},
T.~Ackernley$^{58}$\lhcborcid{0000-0002-5951-3498},
B.~Adeva$^{44}$\lhcborcid{0000-0001-9756-3712},
M.~Adinolfi$^{52}$\lhcborcid{0000-0002-1326-1264},
P.~Adlarson$^{78}$\lhcborcid{0000-0001-6280-3851},
C.~Agapopoulou$^{46}$\lhcborcid{0000-0002-2368-0147},
C.A.~Aidala$^{79}$\lhcborcid{0000-0001-9540-4988}
% ...

{\footnotesize \it

% ...
$^{35}$Nikhef National Institute for Subatomic Physics, Amsterdam, Netherlands\\
% ...
$^{44}$Instituto Galego de F{\'\i}sica de Altas Enerx{\'\i}as (IGFAE), Universidade de Santiago de Compostela, Santiago de Compostela, Spain\\
% ...
$^{46}$European Organization for Nuclear Research (CERN), Geneva, Switzerland\\
% ...
$^{48}$Physik-Institut, Universit{\"a}t Z{\"u}rich, Z{\"u}rich, Switzerland\\
% ...
$^{52}$H.H. Wills Physics Laboratory, University of Bristol, Bristol, United Kingdom\\
% ...
$^{54}$Department of Physics, University of Warwick, Coventry, United Kingdom\\
% ...
$^{57}$School of Physics and Astronomy, University of Glasgow, Glasgow, United Kingdom\\
$^{58}$Oliver Lodge Laboratory, University of Liverpool, Liverpool, United Kingdom\\
% ...
$^{66}$Syracuse University, Syracuse, NY, United States\\
% ...
$^{78}$Department of Physics and Astronomy, Uppsala University, Uppsala, Sweden, associated to $^{57}$\\
$^{79}$University of Michigan, Ann Arbor, MI, United States, associated to $^{66}$\\
% ...
}
\end{flushleft}

\end{document}